\documentclass{emulateapj}
\shorttitle{H$_2$ Emission and Star Formation on Arp 143}
\shortauthors{Beir\~ao et al.}
\usepackage{lscape}
\usepackage{rotating}
\begin{document}

\title{Powerful H$_2$ Emission and Star Formation on the Interacting Galaxy System Arp 143: Observations with {\it Spitzer} and {\it GALEX}}

\author{P. Beir\~ao\altaffilmark{1},
        P. N. Appleton\altaffilmark{2},
        B. R. Brandl\altaffilmark{1}, 
        M. Seibert\altaffilmark{3},
        T. Jarrett\altaffilmark{2}, 
        J. R. Houck\altaffilmark{4}}
\altaffiltext{1}{Sterrewacht Leiden, Leiden University, P. O. Box 9513, 2300 RA Leiden, The Netherlands} 
\altaffiltext{2}{NASA Herschel Science Center, California Institute of Technology, Pasadena, CA 91125}
\altaffiltext{3}{Carnegie Observatories, Pasadena}
\altaffiltext{4}{Astronomy Department, Cornell University, 219 Space Sciences Building, Ithaca, 
                 NY 14853}

\begin{abstract}

We present new mid-infrared ($5 - 35\mu$m) and ultraviolet (1539 -- 2316
\AA) observations of the interacting galaxy system Arp 143 (NGC 2444/2445)
from the Spitzer Space Telescope and GALEX.  In this system, the central nucleus 
of NGC 2445 is surrounded by knots of massive
star-formation in a ring-like structure. We find unusually strong
emission from warm H$_2$ associated with an expanding shock wave between
the nucleus and the western knots.  At this ridge, the flux ratio between
H$_2$ and PAH emission is nearly ten times higher than in the nucleus. 
Arp 143 is one of the most extreme cases known in that regard.
From our multi-wavelength data we derive a narrow age range of the
star-forming knots between 2 Myr and 7.5 Myr, suggesting that the ring of
knots was formed almost simultaneously in response to the shock wave
traced by the H$_2$ emission.  However,  the knots can be further
subdivided in two age groups: those with an age of 2--4 Myr (knots A, C,
E, and F), which are associated with $8\mu$m emission from PAHs, and those
with an age of 7-8 Myr (knots D and G), which show little or no $8\mu$m
emission shells surrounding them.  We attribute this finding to an ageing
effect of the massive clusters which, after about 6 Myr, no longer excite
the PAHs surrounding the knots.

\end{abstract}

\section{Introduction}
\label{intro}

\begin{figure}
\plotone{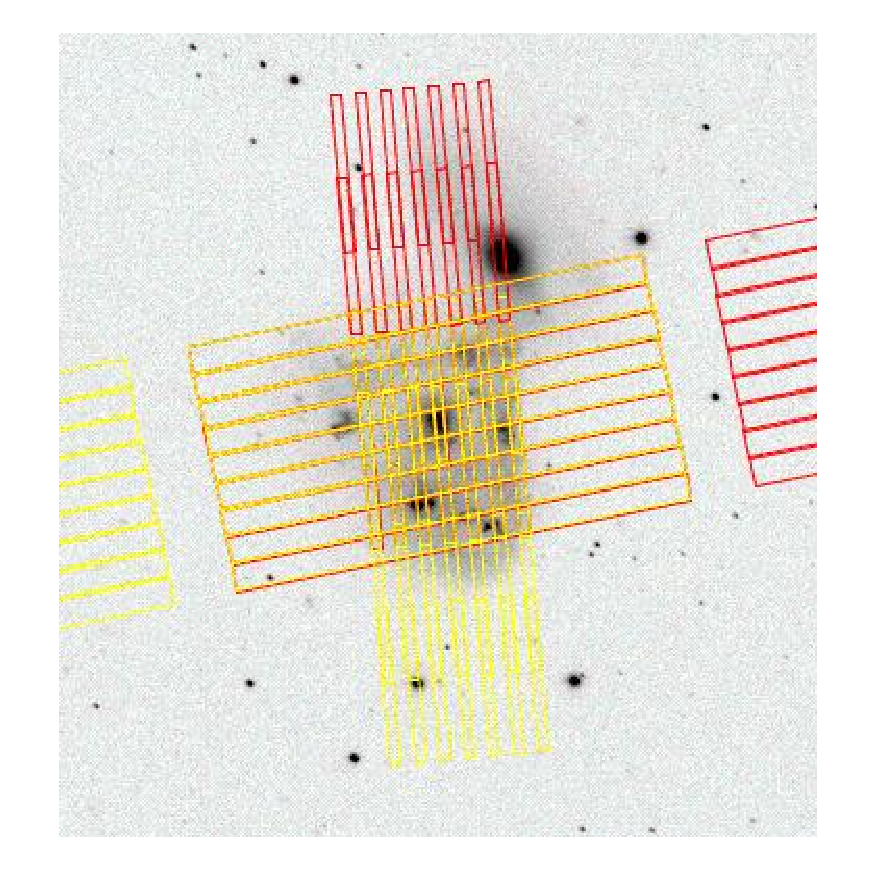}
\caption{Overlay of the SL (sparse) and LL (complete) maps on a V-band image of Arp 143. North is up.}
\label{slits}
\end{figure} 

The direct study of warm molecular hydrogen emission from galaxies has
been improving considerably through the advent of more sensitive space-bourne
spectrometers on the {\it Infrared Space Observatory} (e. g. \citet{rigop02,lutz03}) and more recently on the {\it Spitzer} Space
Telescope\footnote{This work is based, in part, on observations made with the  
Spitzer Space Telescope, which is operated by the Jet Propulsion  
Laboratory, California Institute of Technology under a contract with  
NASA. Support for this work was provided by NASA through a GO2 award  
issued by JPL/Caltech.}. With the Infrared Spectrograph (IRS\footnote{The IRS was a collaborative venture between Cornell University and Ball Aerospace Corporation funded by NASA through the Jet Propulsion Laboratory and the Ames Research Center.}) \citep{houck04} on {\it Spitzer} \citep{werner04} it has been possible to improve upon systematic studies of the
mid-IR pure-rotational lines of molecular hydrogen in a variety of
environments, from nearby normal galaxies \citep{roussel07}, 
to galaxies in more extreme environments, like groups and clusters
\citep{ogle07, egami06, ferland08} and ULIRGS
\citep{armus06, higdon05} (see also a review on molecules by
\citet{omont07}).  

Recently unusually strong mid-IR H$_2$ lines have been
discovered in the giant shock-wave structure in Stephan's Quintet
\citep{apple06}, and in over a dozen low-luminosity radio
galaxies \citep{ogle07,ogle08}, where H$_2$ are often the strongest in mid-infrared wavelengths. Strong H$_2$
equivalent widths were also found near the ``overlap'' region in the
Antenna interacting galaxy \citep{haas05}. These systems have extremely large H$_2$ equivalent widths and
line luminosities (L$_{H_2}> 10^{41-42}$ ergs/s), and relatively
low star formation rates. 
Another defining characteristic is the large ratio of H$_2$ luminosity to PAH
line strength, and the large (0.1 to 30\%) fraction of H$_2$ luminosity to
the total bolometric luminosity of the objects. 
Even larger H$_2$ line luminosities have been found associated with
several massive galaxies in X-ray clusters (e. g. \citet{egami06}). This new class of powerful H$_2$ emitting galaxy appears to be
shock excited, and models of Stephan's Quintet, where a large-scale
shock is strongly implicated suggest that a
significant fraction of the bulk kinetic energy in the shock must be
funnelled into the H$_2$ line in order to explain the results \citep{boula08,guillard08}. In this context, it is of
considerable interest to find other examples of the same kind of
emission in nearby systems. This paper describes strong H$_2$ emission
from the strongly interacting galaxy pair Arp~143.

Arp~143 is an interacting pair of galaxies (NGC 2444/2445) with many
of the necessary ingredients for a study of shocks generated by galaxy
collisions. The southern gas-rich component of the pair, NGC 2445, is
notable for the ring of stellar clusters and HII regions, whereas the
SO companion is devoid of activity. It is possible that NGC 2445
shares some similarities with collisional ring galaxies (like the
Cartwheel ring), because the powerful star-forming knots lie along a
crescent-shaped wave of HI emission \citep{apple92}, and there is
kinematic evidence that the structure is expanding through the
disk \citep{higdon97}. Early unpublished spectroscopy by
\citep{jeske86} and optical and near-IR photometry of the knots
\citep{apple92} suggest the star clusters are quite young
$<$30-60Myrs, much younger than the dynamical age of the expanding HI
wave--thus providing evidence that they are triggered by a collision
with NGC 2444. \citet{jeske86} estimated sub-solar (LMC-like)
metallicity in the disk knots in the range 12+log[O/H]~=~8.56-8.77,
similar to other known collisional ring galaxies (see Bransford et
al. 1996). The nucleus has mildly super-solar values of
12+log[O/H]~=~9.18--similar to nuclear starbursts.

However, this simple picture is complicated by the unsolved mystery of
the large 150kpc-long HI plume \citep{apple87} which extends just to
the north of NGC 2444. The existence of such a long plume implies that
the two galaxies involved in the collision have had a previous major
tidal encounter in the past \citep{hibbard99,hibbard00}. Perhaps the simplest
interpretation of the system is that NGC 2444 may have experienced an
initial close passage with NGC 2445 $\sim$100~Myrs ago, stripping HI
into the plume from NGC 2444. After an initial non-pentrating
encounter, the main stellar body of NGC 2444 has returned to collide
recently with the disk of NGC 2445 in a manner similar to collisional
ring galaxies \citep{apple96}.


In this paper we will present new multi-wavelength imaging and
spectroscopic data of Arp~143, another example of a galaxy with very
strong Mid-IR H$_2$ emission. Our goal is to connect the recent
dynamical events in Arp~143 with the shock and star formation history
of the galaxy.  The paper will use {\it Spitzer}, {\it GALEX\footnote{GALEX (Galaxy Evolution Explorer) is a NASA small explorer
launched in 2003 April. We gratefully acknowledge
NASA's support under Guest Investigator program no. 45.}}, and new
ground based optical and near-IR images as well as mid-IR spectroscopy
to explore the physical conditions in Arp~143. This will include the
study of the dust and HII region properties, ages and luminosities of
star formation regions and the excitation conditions of the warm H$_2$
gas.  After a description of the images and spectra in \S 2, 
we will present in \S 3 the multiwavelength photometry and the spectral
analysis. In \S 4, the discussion of the results will follow two main
directions: the analysis of the shock region using mainly results from the
spectral analysis, and the star formation history of the knots based on an interpretation of the UV-infrared SEDs of each knot.  We
will assume a distance to Arp~143 of 56.7 Mpc based on its
corrected Virgo-centric velocity of 4142 km/s and an assumed H$_0$=73
km/s/Mpc. At this distance, 1 arcsecs corresponds to 275pc.

\section{Observations and Data Reduction}

\subsection{Spectra}
\label{spectra}

\begin{figure*}
\plotone{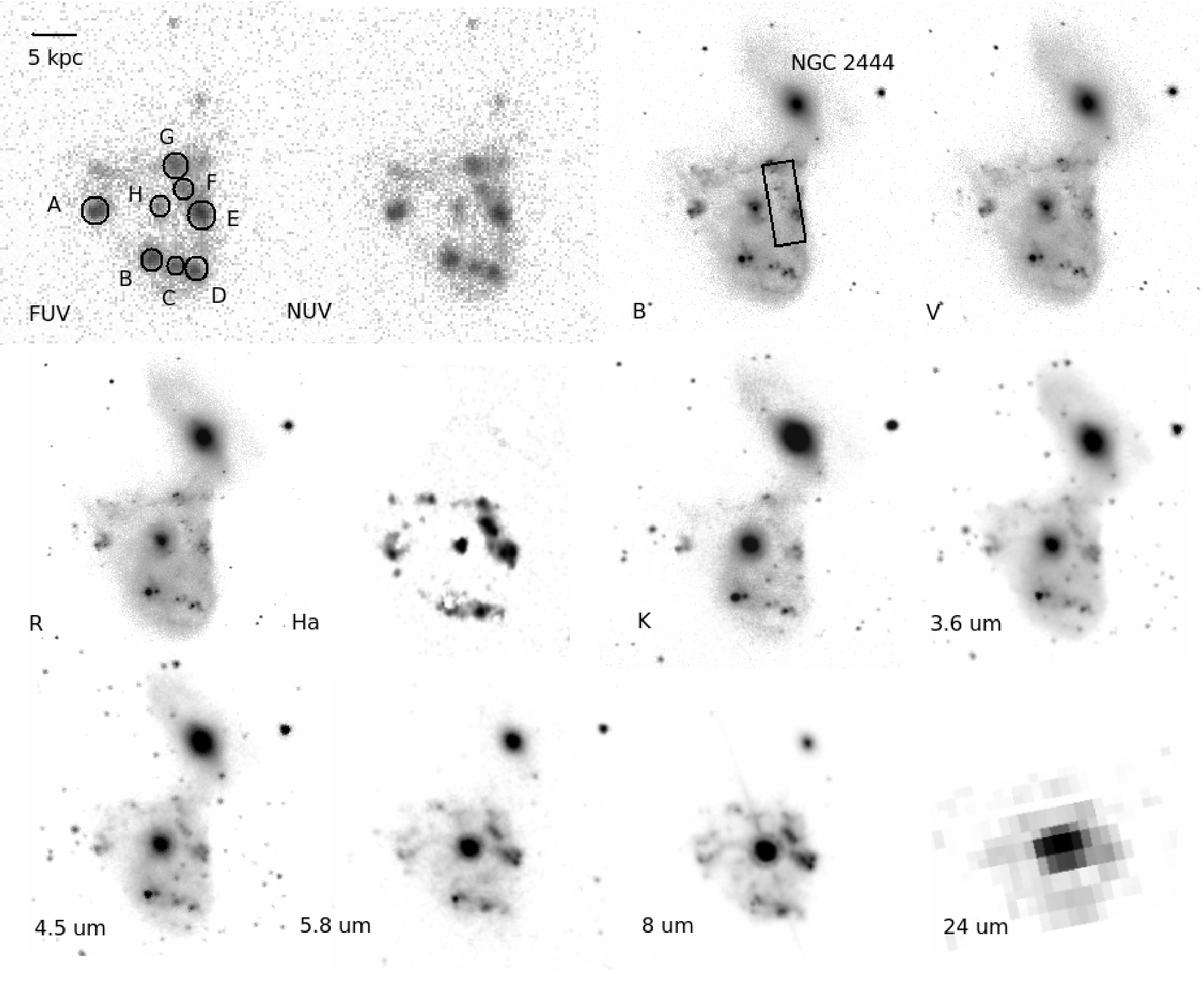}
\caption{Images of Arp 143 in 12 different bands: GALEX FUV, GALEX NUV, B, V, R, H$\alpha$, K, IRAC 3.6$\mu$m, IRAC 4.5$\mu$m, IRAC 5.8$\mu$m, IRAC 8$\mu$m, and IRS LL map at 24 $\mu$m. The different apertures used for photometry are overlayed on the FUV image.}
\label{multiband}
\end{figure*}

The spectra were taken on October 24, 2004 using the IRS
\citep{houck04} on board of the Spitzer Space Telescope. All spectral
data were processed using IRS pipeline version S15. Fig.~\ref{slits}
shows an overlay of the IRS low- and high-resolution slits on a V-band
image.  We have a completely sampled map in the Long-Low (LL; 15 - 37
$\mu$m) spectral mapping mode, consisting of 17 pointings (each with a
32 s exposure): a total LL exposure time of 544 s. A sparsely-sampled
map was obtained with Short-Low (SL; 5 - 14 $\mu$m) which did not
fully sample the galaxy as shown in the figure.  It consisted of 30
pointings with 61 s exposures each: a total SL exposure time 1891
s. Since the LL and SL observations are composed of two sub-slits,
each observed separately, convenient off-source ``background''
exposures were automatically obtained during the observations. These
observations were selected to be free of line emission and were
subtracted from each on-source exposure.

High-resolution spectra of the nucleus were taken in two nods with two
31 s exposures each for short and long wavelengths (SH and LH),
resulting in a total integration time of 248 s. For each module we
averaged the exposures for both nod positions. After the removal of bad pixels, we extracted the spectra
for each module, using SMART\footnote{SMART was developed by the IRS Team at Cornell University and is available through the Spitzer Science Center at Caltech.} \citep{higdon04}, a tool for the extraction and analysis of Spitzer-IRS spectra, written in IDL. 
For the high-resolution spectra we did not subtract a
background since there was no suitable ``sky'' spectrum taken. These observations were made early-on in the Spitzer mission before it was realized that obtaining a ``sky" observation away for the target was useful for rogue-pixel removal (especially LH). Therefore, our SH and LH observations contain low-level zodiacal light in addition to the continuum from the galaxy. This does not affect the measurements of line fluxes (see later) but potentially can affect the measurement of the absolute continuum. The source was sufficiently bright that rogue-pixel removal was not a large issue in the case.

Extraction of all the in-target low-resolution spectra was performed
using CUBISM \citep{smith07b}, an IDL tool for the construction and
analysis of spatially resolved spectral cubes using {\it Spitzer/IRS}
spectra. Bad pixels in the BCD images were manually flagged in each
cube pixel, and then automatically discarded when rebuilding the cube.

\subsection{Images}
\label{images}

Fig.~\ref{multiband} shows images of Arp 143 at 12 different wavelengths: far- ($\lambda~154 $nm) and near-UV ($\lambda~232 $nm) continuum; optical B- V-, and R-band continuum; H$_{\alpha}$ emission; K-band, the mid-infrared $3.6\mu$m and $4.5 \mu$m bands, sensitive to the emission of the red stellar population; the mid-infrared $5.8 \mu$m and $8.0 \mu$m bands, specially sensitive to PAH and dust emission; and a $24\mu$m continuum image, built from a IRS LL spectral map. In the FUV image we show the apertures chosen for photometry, matching the main star forming knots visible on FUV, marked from A to G, and the nucleus. Each image was obtained as follows.

\begin{figure}
\plotone{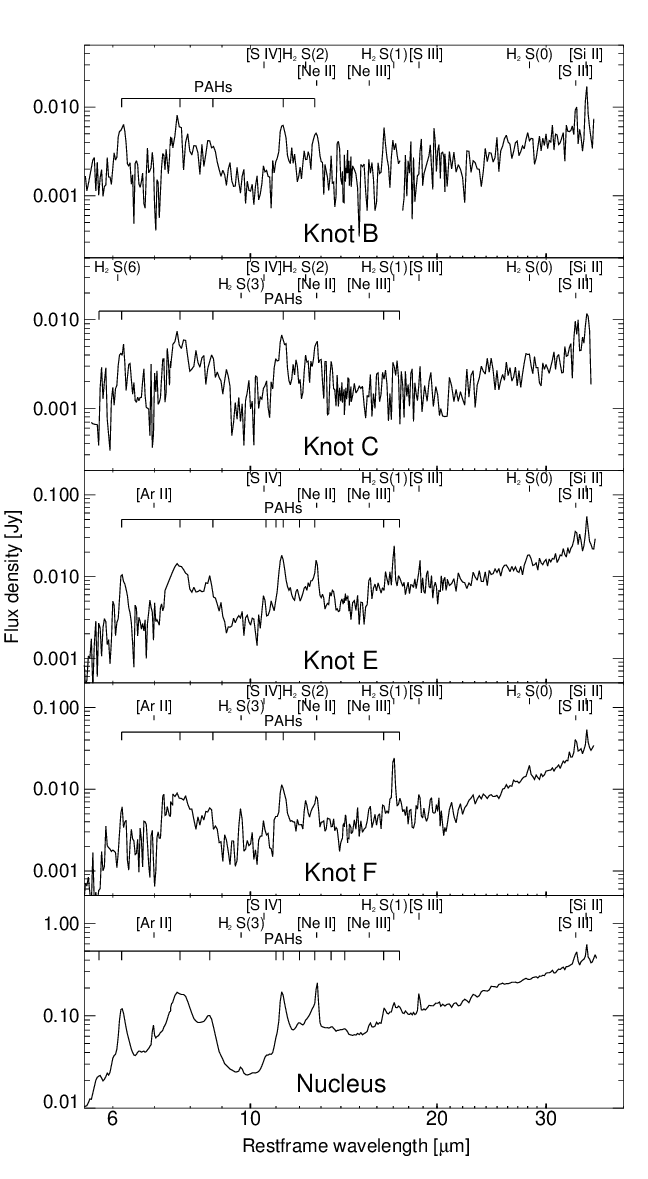}
\caption{Low resolution spectra of six regions coinciding with the selected knots.}
\label{lores}
\end{figure}  

\subsubsection{FUV and NUV}
\label{uv}

Ultraviolet images of Arp 143 were obtained on January 28, 2005 (FUV) and December 26, 2006 (NUV) using the Galaxy Evolution Explorer (GALEX) satellite \citep{martin05}. The galaxy was imaged in the FUV and NUV bands, covering the wavelengths $1344 - 1786 \AA$ and $1771 - 2831 \AA$. The total integration time was 3040 seconds for the FUV image and 219 seconds for the NUV. GALEX uses two 65 mm diameter, microchannel plate detectors, producing circular images of the sky with $1.2\arcdeg$ diameter at $5\arcsec$ resolution. The GALEX image was reduced and calibrated through the GALEX pipeline.The fluxes were converted to magnitudes on the AB system \citep{oke90}.
The data was constructed with version 6 of the GALEX
pipeline. We coadded two All Sky imaging visits to obtain the 219 seconds of exposure in the NUV image. 

\subsubsection{Optical and Near-Infrared}
\label{optical}

We obtained optical images on February 3, 2003 using the Palomar $60\arcsec$ telescope. The observations were made with a FOV of  $12\arcmin.9\times12\arcmin.9$ ($2048 \times 2048$ pixel) CCD imaging system with an RCA chip. The pixel scale is $0\arcsec.38$/pixel The data were taken in three filters, Johnson B, Johnson V, and Johnson R, and exposure times of 500 seconds for each filter. The night was moonless and transparent and the seeing was $0\arcsec.7-1\arcsec.0$. We used the star RU 152 for calibration. The H$\alpha$ image was published in \citet{romano08} and was taken in the Guillermo-Haro 2.1 m Telescope using a narrowband filter centered at $6635\AA$ (FWHM $\sim97\AA$), which also contains the [NII]$\lambda$6583 line. We corrected the H$\alpha$ flux for [NII] by assuming H$\alpha$/[NII]$\sim3$, typical of HII regions \citep{oster89}. The  near-infrared images were obtained with good ($<1\arcsec$)  seeing  through a  K$_s$ ($\lambda$ = 2.15$\mu$m) filter using the  
WIRC camera on the Palomar $200\arcsec$ telescope in November 30 2004.   WIRC  
is a $2048\times 2048$ pixel wide-field $8.5 \times 8.5 \arcmin$ HgCdTe camera  
with a pixel scale of $0.25\arcsec$/pix operated by Caltech.

\subsubsection{Mid-Infrared}
\label{midir}

Arp 143 was imaged with the Infrared Array Camera (IRAC, \citet{fazio04}) on \textit{Spitzer} at 3.6, 4.5, 5.8, and 8.0 $\mu$m on March 28, 2005. The IRAC detectors consist of two $256 \times 256$ square pixel arrays with a pixel size of $1\arcsec22$ resulting in a total field of view of $5\arcsec.2 \times 5\arcsec.2$. For each of the four channels, 47 12 s exposures were taken, with a total integration time of 564 seconds for each channel. The data were reduced using standard procedures (i.e., dark-current subtraction, cosmic ray removal, non-linearity correction, flat-fielding and mosaicing) using pipeline version 14.0 of the Spitzer Science Center.

\begin{figure}
\plotone{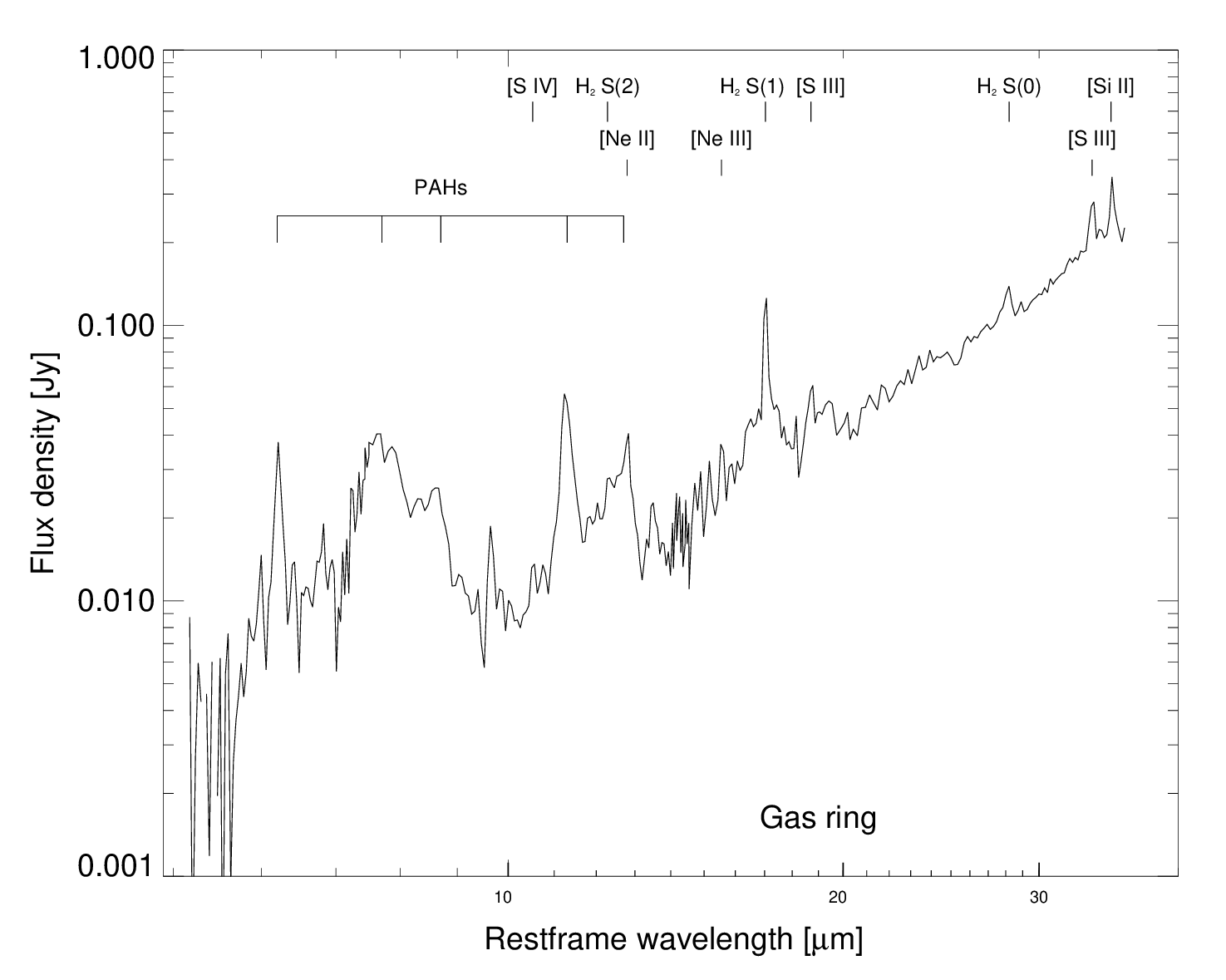}
\caption{Low resolution spectrum of the gas ring in NGC 2445.}
\label{ridge}
\end{figure}

\begin{figure}
\plotone{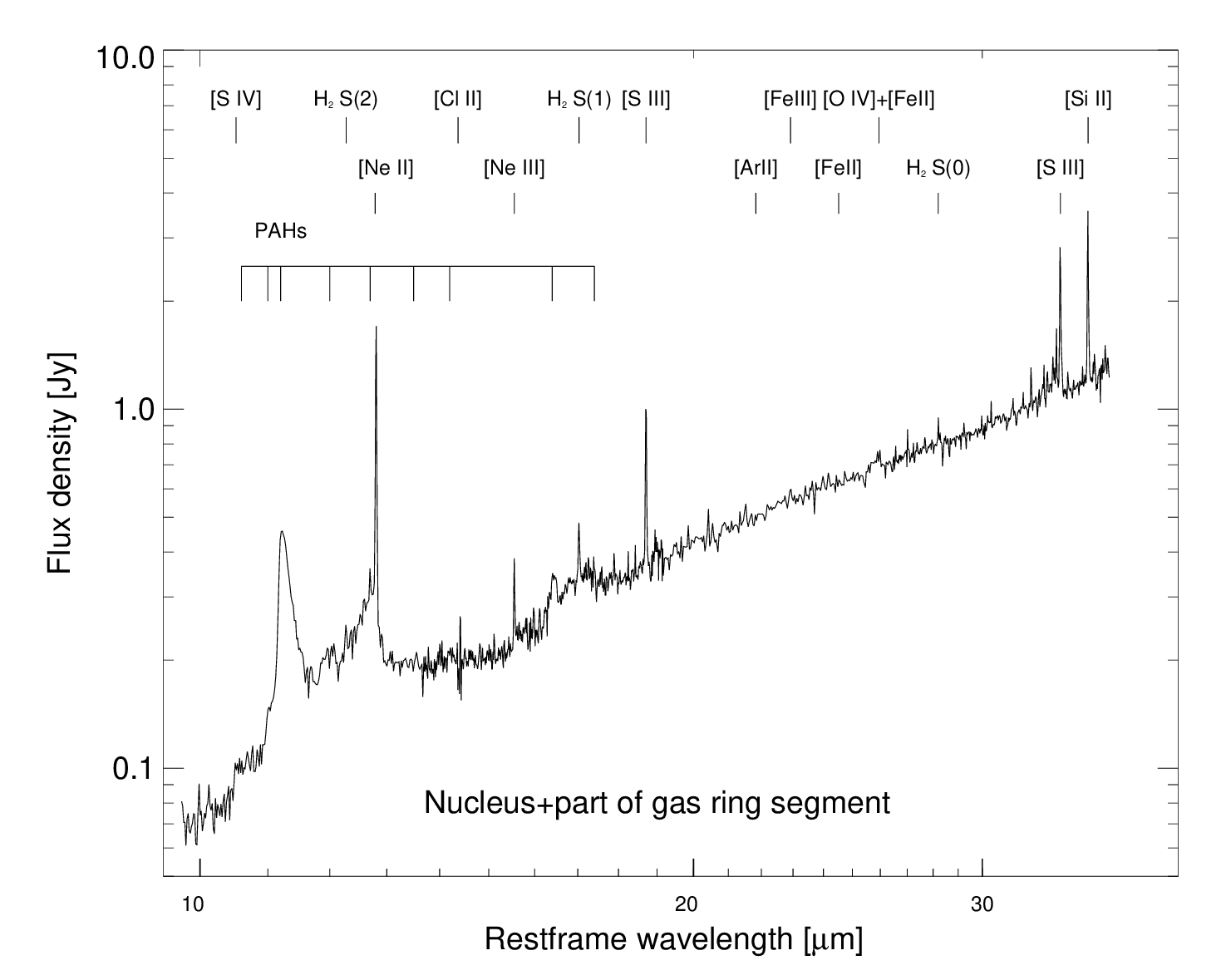}
\caption{Full high resolution spectrum of the nucleus of NGC 2445. The SH part was scaled up by 1.8 to match the LH spectrum.}
\label{hires}
\end{figure}

\section{Analysis}

In Fig.~\ref{lores} are five low-resolution spectra of
$21\arcsec\times21\arcsec$ regions coinciding with knots B, C, E,
F, and the nucleus. In these spectra we can see broad features attributed to
Polycyclic Aromatic Hydrocarbons (PAHs), such as $6.2\mu$m, $7.7\mu$m,
$8.6\mu$m, $11.3\mu$m, $12.6\mu$m features. Also noticeable is the
distinctive PAH $17\mu$m complex. In the regions E and F, the $H_2$
lines at $9.7\mu$m, $12.2\mu$m, $17\mu$m, and $28.2\mu$m are
especially strong. The ionic lines observed are [ArII] at
$6.9\mu$m, [NeII] at $12.8\mu$m, [NeIII] at $15.6\mu$m, [SIII] at
$18.8\mu$m and $33.5\mu$m, and [SiII] at $34.8\mu$m. The spectrum of
knot D is extremely noisy, and therefore it is not shown in Fig.~\ref{lores}. 
This means that there is no continuum flux detected at this wavelength. However, the PAH features are detected above the
sensitivity level. The peak of the $6.2\mu$m PAH feature, for example,
is at 4 mJy, well above the $3\sigma$ sensitivity of the SL slit at
this wavelength, which is 0.6 mJy.
In Fig.~\ref{ridge} is a low resolution spectrum from the region where a ring of HI emission was observed by \citet{apple92,higdon97}, which is delimited in the B-band image in Fig.~\ref{multiband}. This spectrum is very similar to the spectra of knots E and F, but it has some differences. It is even more dominated by distinctive H$_2$ lines, especially the H$_2$ S(1) line at $17\mu$m, although PAH features can also be seen, especially at $11.3\mu$m. The ionic lines are not so prominent, as the spectrum was taken in a region not dominated by star forming clusters.

\subsection{H$_2$ lines}
\label{h2lines}

\begin{figure*}
\epsscale{1.05}
\plotone{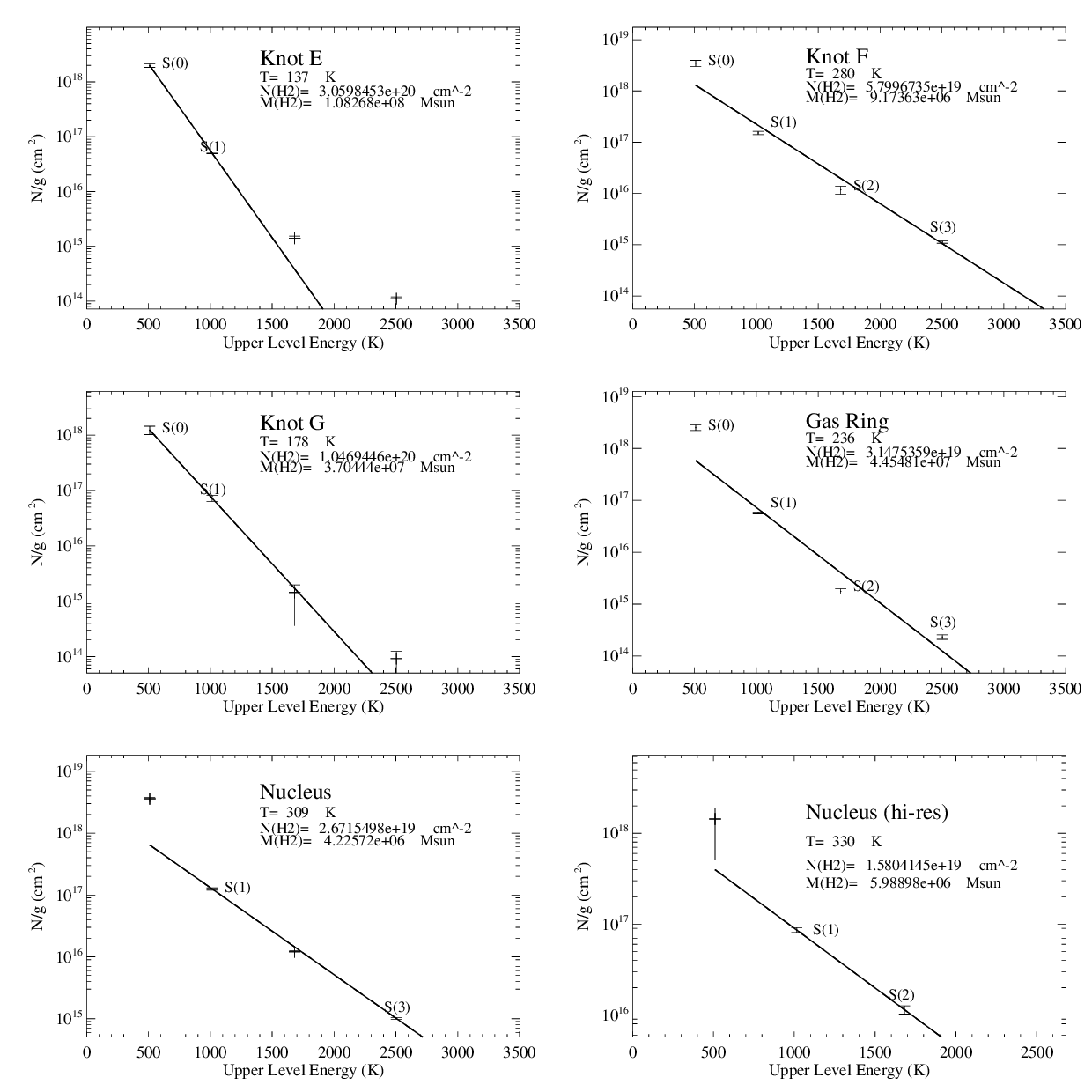}
\caption{H$_2$ excitation diagrams for knots E, F, G, the nucleus, and the gas ring . The plots are labelled with the fitting temperatures (T), column densities N(H$_2$) and molecular gas masses M(H$_2$) in units of K, cm$^{-2}$, and M$_{\odot}$ respectively. The points that are not labeled are upper limits and are not included on the fit.}
\label{h2diag}
\end{figure*}

Using the pure rotational H$_2$ lines we can probe the physical conditions of the warm molecular hydrogen in the star forming knots. The pure rotational lines originate from the warm (T $>100$ K) gas. We measured the H$_2$ rotational line fluxes using SMART, and derived their excitation temperatures, column densities, and masses of their temperature components. Fig.~\ref{h2diag} shows the excitation diagrams of the detected H$_2$ transitions and upper limits for the knots E, F, G, the nucleus, and the ring. Each excitation diagram consist on a plot of the natural logarithm of the column density $N$ divided by
the statistical weight $g$ in the upper level of each transition
against the upper level excitation temperature $T_{ex}$. The column density follows from the Boltzmann equation:

\begin{equation}
\frac{N_i}{N}=\frac{g(i)}{Z(T_{ex})}\times exp\left(-\frac{T_i}{T_{ex}}\right)
\end{equation}

where N(i) is the molecular column density of the ith transition, $N$ is the total column density of H$_2$, g(i) is the statistical weight for the ith
transition, and $Z(T_{ex})$ is the partition function at the excitation temperature $T_{ex}$. The values of g for odd and even transitions are different because of the ortho and para transitions. The H$_2$ line fluxes, and the molecular gas temperatures, column densities and masses are listed in Table~\ref{tabh2}. 

The excitation temperature of the line-emitting gas is the reciprocal of the slope of the excitation diagram, and corresponds to the kinetic temperature in local thermodynamic equilibrium (LTE). However, the nonlinear decline of log (N/g) with upper-level energy, commonly seen in shocks within the Galaxy, as well as in external galaxies \citep{lutz03,rigop02}, is an indication that no single-temperature LTE model fits these data.  

The lowest temperatures are measured in knots E and G, using S(0) and S(1). The temperatures for knots F and H were derived from higher excitation transitions, S(1) and S(2), as S(0) was not detected. The measured temperatures are 137 K for knot E, 280 K for knot F, 178 K for knot G, and 309 K for the nucleus. We can see that the molecular gas is hotter in knot F (280 K) and in the nucleus (309 K), but denser in knot E, where most of the warm molecular gas is concentrated. It is important to note that the temperatures for knots E and G were derived using only the S(0) and S(1) transitions, whereas higher transitions were used in other knots resulting in a higher average temperature of the gas. The derived excitation temperature is also dependent on the ortho-to-para ratio for H$_2$. During the fit to a multi-temperature model through the preferred data points,
we constrain the warm gas by the S(0)/S(2) [para] ratio and
allowing the higher order S(3) and S(5) [ortho] transitions to provide
a very rough guide to the temperature of a hotter component, resulting in ortho-to-para ratios of 2.2 for knot E, 2.7 for knot G, and $\sim 3$ for knot F, the nucleus, and the ridge. These ratios agree with LTE models \citep{burton92}. The various masses are derived multiplying the column density by the physical area of the aperture corresponding to the knot. 

\subsection{PAH features}

We used PAHFIT \citep{smith07} to measure the PAH emission features for each of the knots. Some PAH features are decomposed by PAHFIT into gaussian components, such as the $7.7\mu$m and $11.3\mu$m features, and merged for the calculation of their total flux and EWs. The results of the PAH feature fits are listed in Table~\ref{tabpah}. Most of the PAH features lay on the SL part of the spectrum, meaning that the sparse map only has a 2 pixel wide coverage of the knots. We therefore make an extrapolation of the results taken with these observations to the total area of the knots.

The overwhelming majority of the PAH emission comes from the nucleus, which is also the region with the most emission at $8\mu$m (Fig.~\ref{multiband}). From the star forming ring, Knots A, C, E, and F have PAH associated with the star clusters. Knot E is the one with the highest PAH emission, and this is also expected from the $8\mu$m IRAC image in Fig.~\ref{multiband}). 

\subsection{Ionic lines}

In Table~\ref{tablines} we present the fluxes of the main ionized gas lines in the low resolution spectra, as measured using SMART. The errors listed reflect the S/N ratio at the wavelengths where the lines were measured. With these fluxes we calculated the ionic line ratios [NeIII]/[NeII], [SIII]18.7/[SIII]33.5 $\mu$m, and also [SiII]/[SIII]. 

The [NeIII]/[NeII] rate, because it is a rate between two ionization states of the same element, it is sensitive to the effective temperature of the ionizing stars, and therefore can be used as a measure of the hardness of the radiation field. This ratio varies typically between 0.05 and 1 in starburst galaxies and HII regions in normal galaxies, and is typically greater than 1 in dwarf galaxies \citep{thornley00}. We could only measure the [NeIII]/[NeII] ratio for knots E, F, and for the nucleus, and the results are 0.52, 0.67, and 0.08. The ratios for knots E and F are compatible with very young massive clusters, and are similar to those found in very young clusters in other interacting systems such as the Antennae, where star forming regions with [NeIII]/[NeII] ratios between 0.30-0.73 are found \citet{snijders07b}. The ratio for the nucleus is much smaller than the ratios for the ring knots and are similar to the ratios found in the nuclei of starburst galaxies, like NGC 253 (0.07; \citet{devost04}). 

The [SIII]18.7/[SIII]33.5 $\mu$m ratio, as it is a ratio between two lines of the same ionization state, but with different critical densities for collisions with electrons \citep{oster89}. Therefore it is sensitive only to the density of the ionized ISM. The [SIII]33.5$\mu$m line was detected in all knots, but the [SIII]18.7$\mu$m was detected only in knots E, F, G, and in the nucleus. As seen in Table~\ref{tablines}, the values for [SIII]18.7/[SIII]33.5 $\mu$m ratio for these knots vary between 0.50 in knots E and F and 1.34 in knot G although with a large measurement error for the last case.
 
The [SiII]/[SIII] line ratio is used as a starburst/AGN diagnostic \citep{dale06}. This is because the [Si II]34.82 $\mu$m line is a significant coolant of X-ray ionized regions
or dense photodissociation regions \citep{hollen99}, more commonly associated with AGN, whereas the [S III]33.48 $\mu$m
line is a strong marker of H II regions. The typical value found in AGNs is around 3, whereas for nuclear starbursts is around 1 \citep{dale06}.
We measured this ratio for all knots, varying from 0.82$\pm$0.22 in the nucleus to 1.97$\pm$0.51. The low signal-to-noise ratio in clusters B, C, and D is reflected in the high errors of the [SiII]/[SIII] measurements, and therefore all the values for this ratio are compatible with the values for nuclear starbursts found in \citep{dale06}.

\subsection{High resolution spectrum of the nucleus}

To study the ISM conditions in the nucleus with greater accuracy we took a high resolution spectrum of the nucleus of NGC 2445, which is presented in Fig. ~\ref{hires}. We scaled up the SH spectrum by multiplying it by a factor of 1.7 to match the LH spectrum flux at $20\mu$m. This factor is much lower than the difference between the SH and LH slits, but that is simply due to the fact that the nucleus is practically a point source. The LH and SH spectra were then joined together,
resulting in a single $10 - 35\mu$m high resolution spectrum of the
nucleus. The main features present are the ionic lines, like [NeII], [NeIII], [SIII], and [SiII]. PAH bands are also visible, as the $11.3\mu$m band, and the $16 - 18\mu$m feature. Also present are the H$_2$ rotational lines. These features allow a physical characterization of the ISM of the nucleus, and they can be used to diagnose electron temperature and density and H$_2$ gas temperature. 
The measured line ratios are listed in Table~\ref{tablines}, along with the lines measured in the same knot using the low resolution spectrum. We can see that the low resolution line fluxes have roughly half the flux as the high resolution lines. This is due to the difference in the extraction apertures. The area of the LH slit is 231 arcsec$^2$, nearly twice the area of the low resolution extraction aperture chosen for the nucleus, which is 104 arcsec$^2$. This gives a ratio of 2.22, whereas the ratio between low and high resolution apertures for both the [NeII] and [NeIII] flux is 2.04, meaning that the surface brightness of the lines decreases with a wider aperture. However, to do a comparison of the continuum fluxes one has to estimate the background flux. We measured the background flux from an off-source SL slit, which gives 116 mJy and 95 mJy for 15$\mu$m and 24$\mu$m, after scaling for aperture sizes. That means that the background accounts for $\sim55\%$ of the high-resolution continuum at 15$\mu$m and $\sim32\%$ of the high-resolution continuum at 24$\mu$m. The ratio of the continuum flux at 15 $\mu$m is thus 1.6, meaning that the surface brightness of the warm dust decreases with the aperture. The ratio for the $24\mu$m continuum is 2.2, meaning that the surface brightness of cooler dust does not vary with slit size.

Comparing the ratios derived from high resolution with the low resolution ratios, we find a remarkable match for the [NeIII]/[NeII] ratio, being 0.08 for both spectra, but for the [SIII]18.7/[SIII]33.5 $\mu$m ratio there are significant differences, being 0.73 for the high resolution spectrum and 0.49 for the low resolution spectrum. This reflects the different flux ratios between high and low resolution apertures: [SIII]18.7$\mu$m has a flux ratio of 2.38$\pm$0.16 thus maintaining its surface brightness, whereas [SIII]33.5$\mu$m has a flux ratio of 1.61$\pm$0.09 thus decreasing its surface brightness. This difference possibly reflects a decrease of the ionized gas density as the distance to the nucleus increases. Note that the LH slit could be contaminated by emission form the ``shocked'' gas region, and thus it can affect the line ratios that use lines [SiII] and [SIII]33.5$\mu$m and overestimate the nuclear H$_2$ component.

\subsection{Photometry}
\label{photometry}

Photometry of each knot was performed using \textit{aper}, an IDL tool from the IDL Astronomy Library. We used the FUV GALEX image to set an aperture size for each knot, as seen in Fig.~\ref{multiband}. We used circular apertures of $4\arcsec.5$ for knots C, F, and the nucleus; $6\arcsec$ for knots B and D; and $7\arcsec.5$ for knots A, F, and G. The sky background was determined using circular apertures of the same size as the ones used to measure each knot. With these apertures, we measured the flux within selected regions inside Arp 143 -- but avoiding the knots -- and averaged the results. We applied extended source corrections for IRAC photometry. These corrections depend on the size of the aperture and do not exceed 6.8\%. The resulting fluxes of each of the knots are listed in Table~\ref{tabfluxes}.

Arp~143 was not observed with the MIPS instrument at 24$\mu$m.  
However, in order to allow us to use some of the well known star- 
formation
indicators (e. g. \citet{calzetti07} ) associated with the knots based  
on MIPS observation, we decided to create pseudo-MIPS 24$\mu$m photometric  
points by extracting 24$\mu$m data from the full-IRS LL map (see  
Fig.~\ref{multiband}--last panel) using the CUBISM-extracted spectra,  and the  
MIPS 24$\mu$m filter response curve.  These values are tabulated in  
Table~\ref{tabfluxes}.

\section{Results}

Our goal is to characterize the physical origin of the H$_2$ line emission, its role on the appearance of the ring of star formation around the nucleus of NGC 2445, describe how the ISM in the knots evolves as they age. To achieve this goal we now analyze the observations described in the past chapter according to the following points: a) the study of the morphology of the system, describing it in the present state, in order to compare it to the present theories on the evolution of ring galaxies; b) the study of $H_2$ excitation region and comparison with predictions, in order to study its physical origin; c) the ISM properties of the knots and its evolution, using the mid-infrared spectra to study the ISM in the knots, and UV to K-band photometry to study their ages; and d) the star formation rates in the knots, and the future of star formation in the system.  

\subsection{Morphology}
\label{morphology}

\begin{figure}
\plotone{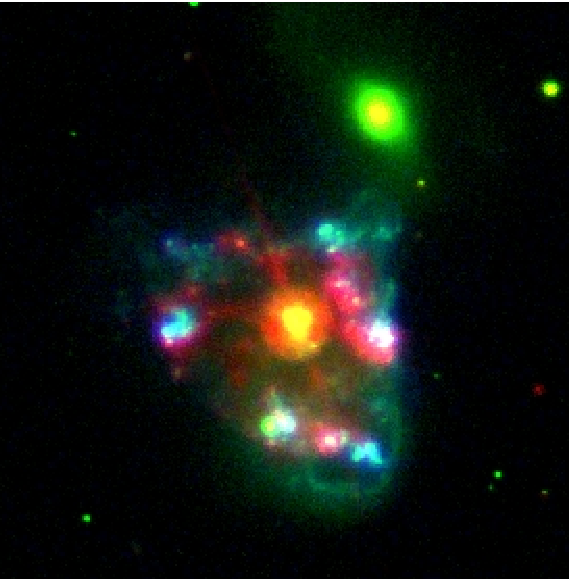}
\caption{Composite of FUV (blue), V-band (green), and 8 $\mu$m (red) images of Arp 143.}
\label{multicolor}
\end{figure} 

In Fig.~\ref{multicolor} we present a composite of FUV (blue), V-band (green), and 8.0 $\mu$m emission (red), which traces emission from hot dust. Most of the luminosity in the V-band comes from the nuclei of the two galaxies, NGC 2444, and NGC 2445, and a ring of star formation knots that surrounds the nucleus of NGC 2445. The nucleus of NGC 2445 dominates the emission in the infrared, indicating the presence of large amounts of warm dust, but it also has a large unobscured population of main sequence and old stars, as it has also significant optical and near infrared components. For a more quantitative comparison, the $8\mu$m flux from from the nucleus is about 5 times the combined $8\mu$m flux from the knots. The knots are located along the outer edge of an HI crescent \citep{apple92}. The knots are composed by several massive young clusters, visible in blue, and some are surrounded by PAH emission, visible in red. There are some mid-infrared ``arms'' connecting the nucleus to some of these regions. These arms are connected to the collisional nature of the ring -- spokes are expected in gas that is collecting downstream of the ring \citep{struck96}. The only other galaxy with similar features reported in the literature is the Cartwheel galaxy \citep{struck96}. 

NGC 2445 is about 20 kpc across. Some peculiarities reveal a connection between the massive star clusters and the dust, like the easternmost knot at about 12 kpc from the nucleus, identified in the FUV image in Fig.~\ref{multiband} as knot A. In Fig.~\ref{multicolor} this FUV emitting knot seems to be surrounded by $8\mu$m mid infrared emission (in red), as it is brighter around the knot and less bright inside a radius of $\sim1$ kpc from the center of the knot. This can be interpreted as a shell of warm dust heated by the massive stars. Other knots have 8 $\mu$m counterparts, like knot B, C, E, and F. For knots C and F, the $8 \mu$m emission can be seen at the same site as the optical cluster, whereas in knot E, the $8 \mu$m is found somewhat offset to the SE. Since the $8\mu$m IRAC band is dominated by PAH emission features, this means that the knots can be divided into two groups: those with associated PAH emission (A, C, E, and F), and ``bare" knots, for which little or no PAH counterpart is observed (B, D, and G). This confirms what is observed in the spectra of Fig.~\ref{lores}, where PAH features are barely observed in knot D. The nucleus of NGC 2444 also shows substantial 8 $\mu$m emission, reflecting substantial massive star formation. The FUV emission comes mainly from knots A, E, and G, showing emission from young massive stars, whereas the IR-bright knot F emits little UV. An even more striking difference between mid-infrared and UV brightness, and Fig.~\ref{multiband} shows it clearly, is seen in the nucleus, the brightest region in 8$\mu$m, with very little UV emission. 

The companion lenticular galaxy NGC 2444 is prominent only in the
optical and K-band, and is rather weak at longer wavelength. This
supports the idea that much of the gas has been swept out of the  
galaxy in the past
and it is dominated by an old inactive stellar population. It is  
possible that a faint``arm" extended
from NGC 2444 behind NGC 2445, but this is hard to prove without  
kinematic data.

\subsection{The Shock Front}
\label{shock}

\begin{figure*}
\plotone{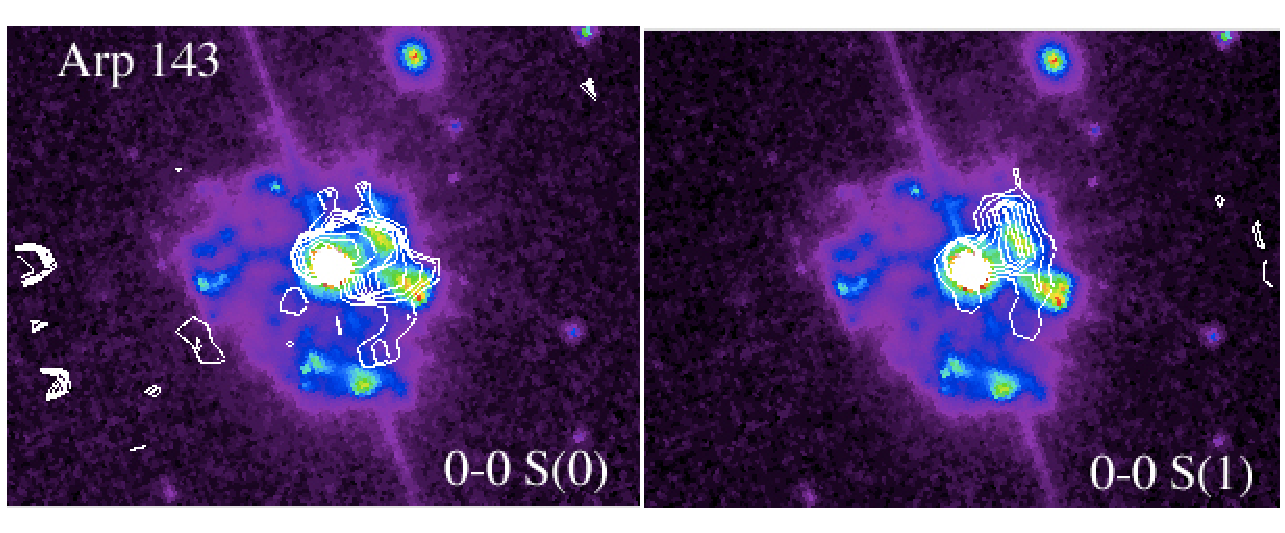}
\caption{Contour maps of the H$_2$ S(0) (left) and H$_2$ S(1) (right) emission in Arp 143, overlayed on an IRAC $8\mu$m image. Contour levels are at 0.5, 0.6, 0.7, 0.8, 0.9, 1, 1.5 , 2, and 3 MJy/sr for H$_2$ S(0) and 1, 1.4, 1.8, 2, 2.5, 3, 4, 6, and 8 MJy/sr for H$_2$ S(1).}
\label{h2fig}
\end{figure*} 

Rotational emission lines of H$_2$ may arise through three different mechanisms:
UV excitation in PDRs surrounding or adjacent to the {\sc HII} regions; shocks that accelerate and modify the gas in a cloud, collisionally exciting the H$_{2}$ molecules; and hard X-ray photons capable of penetrating the molecular clouds and heating large ionizing columns of gas. Several rotational H$_2$ lines were detected in the wavelength range of our observations and they are listed in Table~\ref{tabh2}.  Given that these lines trace the molecular gas at different temperatures, studying the spatial distribution of these excitation lines may give us a clue about the main excitation mechanism.

\begin{figure}
\plotone{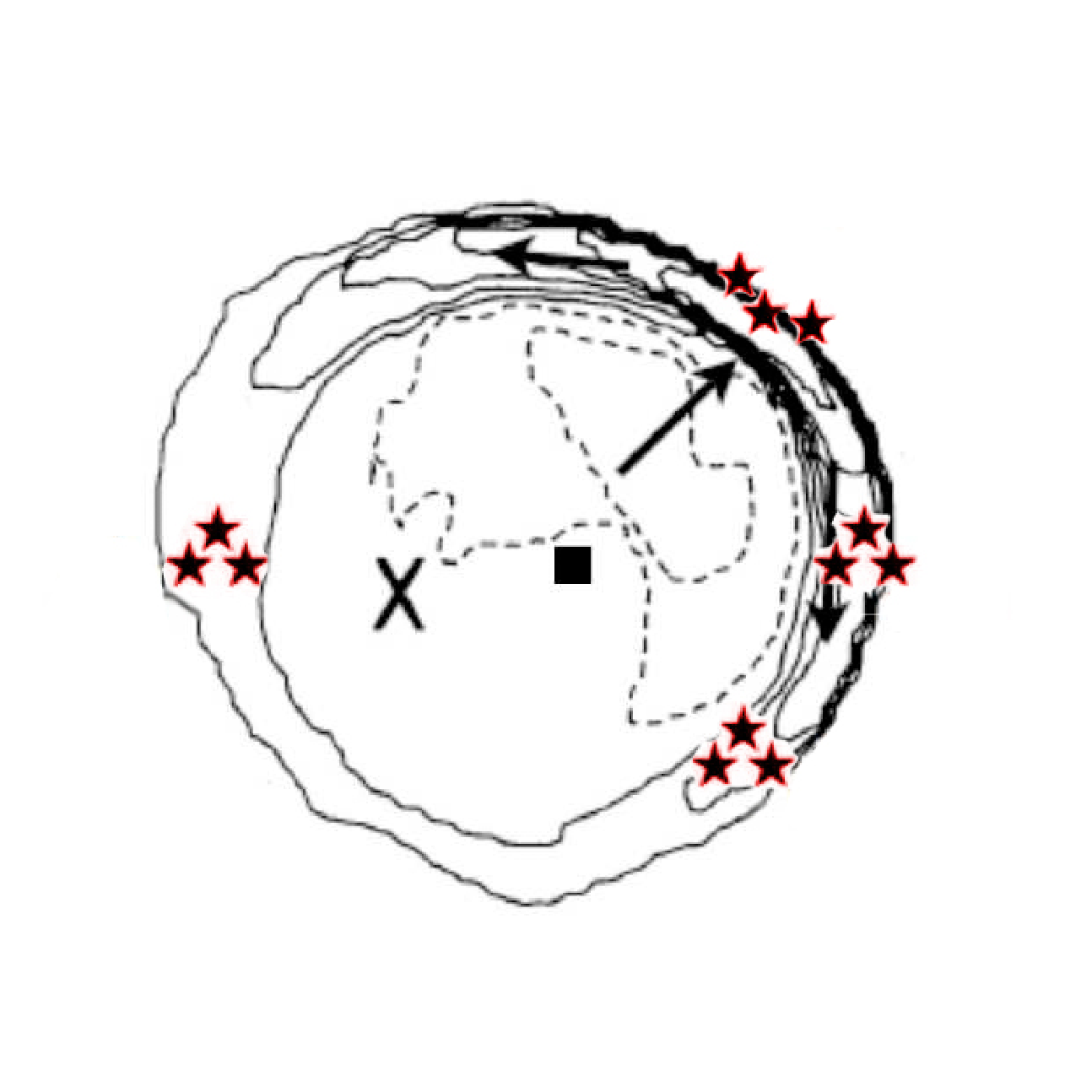}
\caption{Schematic scenario of NGC 2445 where the knots are situated at the leading edge of the density wave. The contours are from the off-center collision model E from Fig. 7 in \citet{apple87}, representing a stage of the expansion process of a density wave. The arrows represent the expansion of the wave. Solid contours indicate elevation above the initial unperturbed disk values and dotted contours represent depressed levels. The nucleus is marked with a square. The site of the collision is marked with a cross.}
\label{cartoon}
\end{figure}

In Fig.~\ref{h2fig} we present contour maps of the H$_2$ S(0) (left)
and H2 S(1) emission (right) in Arp 143, overlayed on
the IRAC 8$\mu$m emission color map. The H$_2$ emission in
Fig.~\ref{h2fig} is concentrated in two clumps--from the nucleus and in a  
crescent-shaped ridge (including knot F) west of the nucleus corresponding to  
the HI-crescent seen
in VLA observations.  The emission also is similar, but not identical to
CO (1-0) interferometry distribution\citep{higdon97}. These latter
observations trace the distribution of cool H$_2$ emission through  
collisional excitation of the CO molecule.
The main difference between the Spitzer (warm) H$_2$ observations  
and the CO (cool H$_2$ tracer)
is that the warm gas seems to better define a crescent-shape. Indeed  
it is precisely here that a shock wave would be expected to heat up the H$_2$, supported by the fact  
that the S(1) map more clearly defines
the crescent shape than the S(0) line, which is dominated by cooler  
excitation. This is compatible with the theoretical scenario illustrated in Fig.~\ref{cartoon}, taken from \citet{apple87}, which represents the density wave formed after an off-center collision between two galaxies (the site of the collision is marked with an X), which expands outwards from the center and also grows in length. It is also worth noting that the area chosen for the spectrum of the gas ring, demarcated in Fig.~\ref{multiband} coincides with the area of the shocked H$_2$ emission. Therefore we can consider this spectrum, in Fig.~\ref{ridge} as representative of the shocked H$_2$ region.

The shape of the H$_2$ emission in Fig.~\ref{h2fig} is promising but by itself does not settle the case for either PDR or shock origin. A way to investigate the origin of the excitation of rotational transitions of the H$_2$ molecule is to make a spatial comparison between the H$_2$ emission and PAH emission. The H$_2$  
emission-line strength should track the PAH emission-line and IR  
continuum strength in the case where the most common H$_2$ excitation mechanism is UV excitation at the PDR interface with star formation regions. However, we see a remarkable transition in  
Arp 143, from the nucleus, where PAH emission is quite strong, to the
non-nuclear H$_2$, which remains strong in the H$_2$ lines, but is very  
weak in PAH  (and continuum) emission. 

\citet{roussel07} calculated the logarithm of the average ratio of the power emitted in the sum of the S(0) to S(2) transitions to the power emitted by the PAH features within the IRAC4 band, for HII nuclei and Seyferts from the SINGS survey. The average ratio for HII nuclei $-2.19\pm0.10$ and for the Seyferts is $-1.80\pm0.34$. We measured these ratios for the knots where H$_2$ lines from S(0) to S(2) have been measured. We take 7.7$\mu$m and  8.6$\mu$m as the PAH features inside the IRAC4 band. For the nucleus of Arp 143, the ratio is -2.26, which puts it in the average of the HII nuclei. However, for knots E, F, and, G, the ratios are -1.22, -0.90, and -0.94, respectively. This puts them well above the averages for HII nuclei and also for Seyferts. For the gas ring region, which largely coincides with the H$_2$ crescent, this ratio rises up to -0.36.  So the difference we observe in the H$_2$/PAH ratio between the nucleus and the ring region in Arp 143 is a strong indication that the H$_2$ emission ridge seen in Fig.~\ref{ridge} is mostly due to shocks. X-ray emission is unlikely to be a strong contribution to the H$_2$ heating, since the  
ratio of the [SiII]/[SIII] lines in Arp 143 (even in the nucleus) suggests an  
insignificant contribution to the excitation of [SiII] from X-rays.  
(X-rays can significantly enhance this ratio--as discussed by \citet{dale06}). This strongly suggests shock-excitation within the crescent-shaped structure. A very strong H$_2$ emission in the absence of strong PAH emission is seen in the shock wave in Stephan's Quintet \citep{apple06} and  
in the spectrum of 3C326 \citep{ogle07}, where strong shocks are implicated. The lack of a correspondingly large continuum is also one of the  
characteristics of the shocked H$_2$ seen in the radio galaxy sample,  
and the Stephan's Quintet, as well as low-excitation emission lines. 

We can interpret the excitation diagrams in Fig.~\ref{h2diag} also in the context of the different origins of the H$_2$ emission. The temperatures derived in the excitation diagrams are compatible with either a PDR origin or a shock origin for the H$_2$ excitation. The H$_2$ temperature derived for the nucleus is higher than the temperature for the knots associated to the H$_2$ ridge, and to the H$_2$ ridge as a whole, still does not reach the level required to rule out PDR origin, and the comparison with the PAH emission is crucial. However, these temperatures are just average gas temperatures, and it is possible the presence of a warmer gas component in the H$_2$ shock region. The column density of the gas is lower in the H$_2$ ridge than in the knots, which is expected, since the warm H$_2$ emission from the ridge is more diffuse, whereas the H$_2$ emission peak coincides with knot F.

\subsection{The Properties of Gas and PAHs in the Knots}
\label{ism}

The knots in the ring of star formation are very young and are expected to have similar characteristics to young massive clusters discovered in starburst galaxies. We use our observations to study the components of the ISM in these knots, which consist of the molecular gas, ionized gas, and dust. 

The properties of the warm molecular gas, as presented in Sect.~\ref{shock}, can be compared with the properties of the cold molecular gas as studied in \citep{higdon97} using CO observations. CO probes molecular gas up to a temperature of 100 K, but provides no information on the "warmer'' gas which might be more directly linked to the source of activity. However, we can estimate what fraction of the total gas mass lies in higher temperatures. The fraction H$_2$ (warm) to H$_2$ (cool) is 4.8\% for knots F and G, 2.6\% for the nucleus, but 64\% for knot E, assuming a CO-H$_2$ conversion factor for the LMC. On average, the fraction of warm to cool H$_2$ amounts to 10\%, which is nearly the same found by \citet{rigop02} for a sample of starburst galaxies.

Forbidden ionic lines can be used as indicators of the conditions of the ionized gas in the knots and the shock region. The ratio [NeIII]/[NeII] is used as a measure of electron temperature of the ionized gas; the [SIII]18.7/[SIII]33.5 $\mu$m ratio is used as a measure of electron density. These ratios are compared to modeled ratios derived using photoionization diagnostics by \citet{snijders07}, assuming Salpeter IMF, stellar mass cutoffs M$_{up}=100 M_{\odot}$ and $M_{low}=0.02 M_{\odot}$ and solar metallicity. 

From the [SIII]18.7/[SIII]33.5 $\mu$m ratio we can estimate the density of the knots for which these lines are measured. This ratio is insensitive to the ionization parameter $Q$ and age. As seen in Fig.~\ref{ratios} (right), for any $Q$ and age, knots E, F, G, and the nucleus have electron densities in the order of $10^2-10^3$ cm$^{-3}$. This is the average electron density encountered in the center of the starburst galaxy M82 \citep{schreiber01}. However Knot E, for example has a diameter of $\sim1$ kpc, about two times the size of the central region of M82 \citep{beirao08}, and it is composed of individual clusters that could have much higher densities.

Fig.~\ref{ratios} in the left shows the evolution of the modeled [NeIII]/[NeII] with age, for a given value of the ionization parameter. Due to difficulties in measuring the [NeII] line in the low resolution spectra, as it is merged with a PAH feature at $12.8\mu$m, we could only measure the ratio for knots E, F, and for the nucleus. We can see that the measured ratios correspond to knot ages of 1 - 4.5 Myr. The difference in the [NeIII]/[NeII] ratio between the nucleus (0.08) an the ring knots (0.5-0.7) means that population of recently formed massive stars in the nucleus is older than the young bursts in the ring, having an estimated age of 5 - 6 Myr. The fit to the low resolution spectrum of the nucleus gives an A$_{\rm V}$ of 0.02, which is too low to have any effect on the ratios, and therefore extinction does not affect the ratios significantly. 

PAH molecules with different physical characteristics produce different bands. The $6.2\mu$m, $7.7\mu$m, and the $8.6\mu$m bands are produced preferentially by ionized PAHs whereas the $11.3\mu$m band is produced primarily by neutral PAH molecules. Therefore, the ratio $11.3/7.7\mu$m may indicate the effect of PAH ionization. We calculate this ratio for all the knots. Given the uncertainties in the PAH flux measurements, we could only derive a reliable $11.3/7.7\mu$m ratio for knots F and H. We cannot see significant differences between the two ratios, at least not bigger than the uncertainties for knot F.


\subsection{The Ages of the Knots}
\label{ages}

With the fluxes listed in Table~\ref{tabfluxes}, we built SEDs of the main knots, in order to further constrain their physical characteristics. To achieve this, we compare the SEDs with synthetic spectra modeled with Starburst99 \citet{leitherer95}. There are several parameters we need to consider when fitting the model SEDs to our observations:

\begin{itemize}
\item The mode of star formation. The star formation may occur continuously (continuous star formation model, or CSF model) or in a single, almost instantaneous, burst (instantaneous burst model, or ISB). We will adopt the ISB model for the star-forming knots, as there is no evidence that star formation has been ongoing for any considerable time in these knots.
\item The stellar initial mass function (IMF). We adopt a Kroupa IMF with an upper-mass cutoff of 100 M$_\odot$ and a lower mass cutoff of 0.02 M$_\odot$.
\item The initial gas metallicity (Z). We assume sub-solar metallicity $Z=0.4Z_{\odot}$, which is similar with the metallicity reported by \citet{jeske86}.
\item The age of the star clusters. This is a free parameter ranging from $0$ to $1\times10^9$ yr.
\item The effect of extinction. We use the Calzetti starburst extinction law \citep{calzetti00} on the synthetic spectra, 
\item Dust emission. 
\end{itemize}

With these conditions, we fitted the synthetic spectra to the measured SEDs of each knot. The fluxes listed in Table~\ref{tabfluxes} were converted to units of ergs$^{-1}$\AA$^{-1}$. We compensated the optical and UV fluxes for a galactic reddening of E(B-V)=0.051, using the extinction maps and laws of \citet{schlegel98} and attenuation curve of \citet{cardelli89}, the latter exclusively for the UV fluxes. Along with a pure stellar synthetic spectrum, the output of Starburst99 includes a nebular emission component as well, which was added to the stellar spectrum for knots C, E, and F. A set of free parameters, such as age, extinction, and continuum flux were adjusted for each knot in order to minimize the $\chi^2$ value. The results are listed in Table~\ref{sed}, as well as the range of parameters within a 95 \% confidence level. Knot B was left out, as knot B has a foreground star that interferes with the measurements. Although the nucleus could be modeled using the above assumptions, the measured optical to UV flux ratio appears always larger than the modeled spectrum. The variation between subsolar metallicity $Z=0.4Z_{\odot}$ and solar metallicity does not modify significantly the parameter values.

\begin{figure*}
\epsscale{1.15}
\plottwo{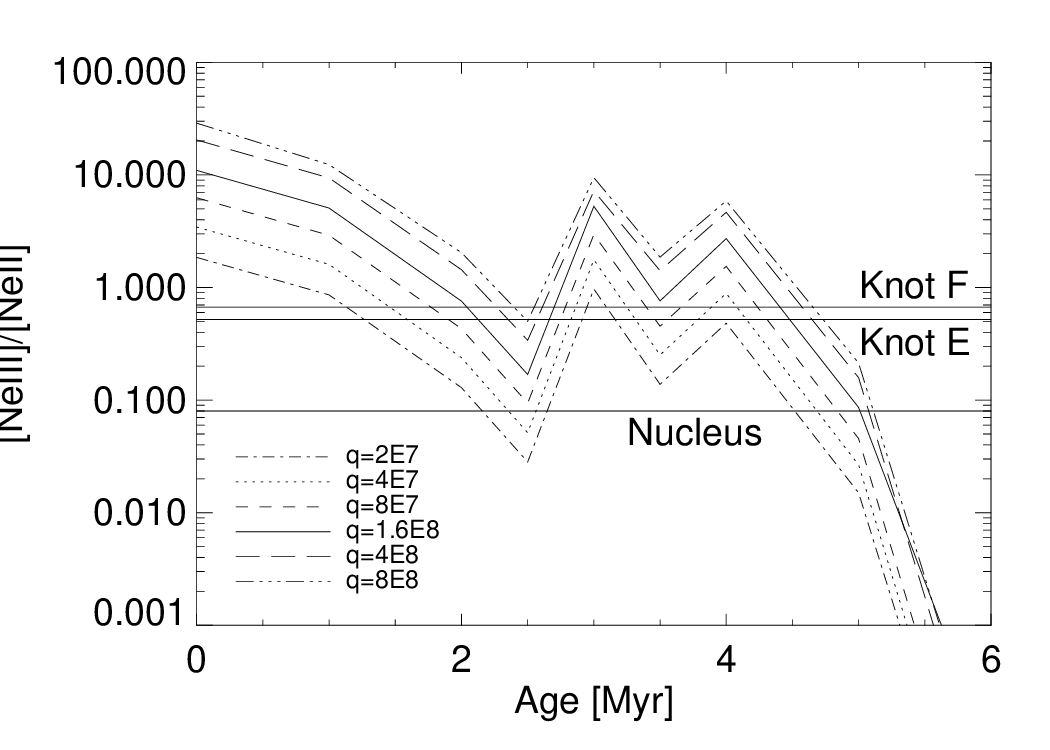}{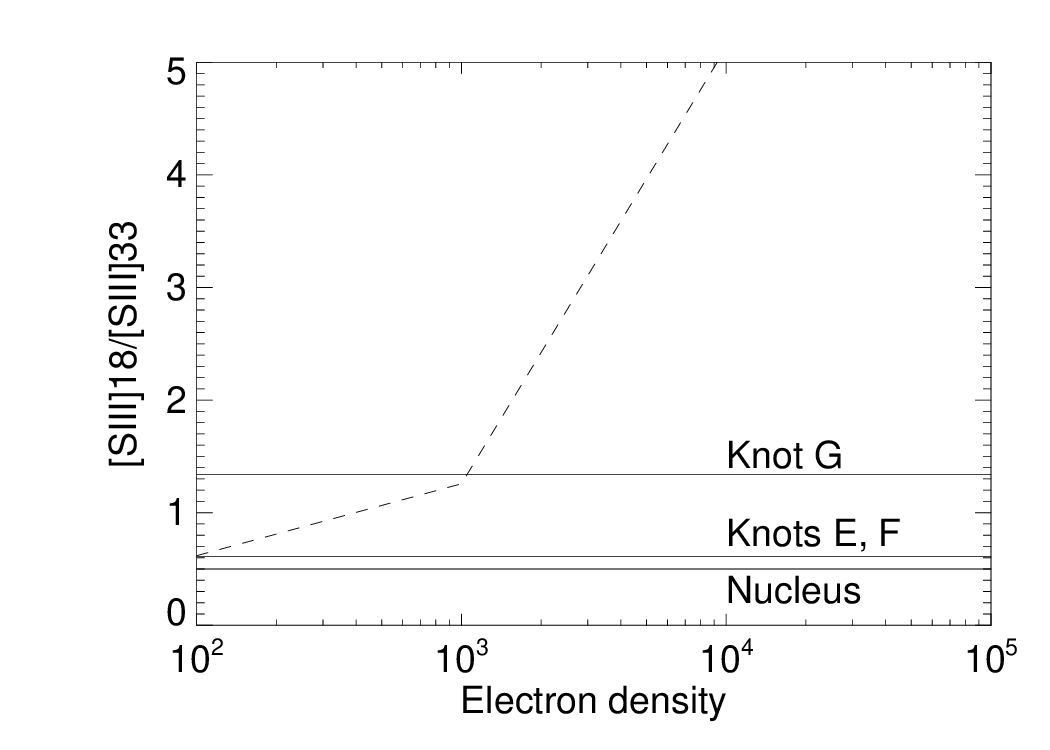}
\caption{\textit{Left}: effect of the cluster age on the [NeIII]/[NeII] ratio. The model curves are computed for a cluster of $10^6M_{\odot}$, assuming Salpeter IMF, $M_{up}=100M_{\odot}$, and $M_{low}=0.2M_{\odot}$. Each curve represents a different ionization parameter, and the solid curve is the one that approaches the value found in M82 by \citet{schreiber01}, log$U=-2.3$. The horizontal lines indicate the [NeIII]/[NeII] for knots E and F, and the nucleus. The first dip in [NeIII]/[NeII] represents the ageing of the stellar population, after which WR stars are produced, increasing the [NeIII]/[NeII] ratio. The dip at 6 Myr occurs as most massive stars die through supernova explosions. \textit{Right}: dependency of [SIII]18.7/[SIII]33.5 $\mu$m ratio on the electron density. The curve is for a single ionization parameter approaching log$U=-2.3$. The horizontal lines indicate [SIII]18.7/[SIII]33.5 for knots E, F, and G, and the nucleus. Knots E, and F have the same value for [SIII]18.7/[SIII]33.5, which is 0.5.}
\label{ratios}
\end{figure*}

From the results of the SED fits in Table~\ref{sed} we can divide the knots in two age groups: knots with ages from 2 - 4 Myr (A, C, E, and F), and knots with ages from 7 - 7.5 Myr (D, and G). Given that the interaction occurred 100 Myr ago \citet{apple87}, this age difference is not very significant in terms of the interaction dynamics. Probably due to a clumpy ISM, some clouds collapsed earlier than others, forming stars a little ahead in time than others. The estimated ages come into agreement with Fig.~\ref{ratios} (left), which predicted the same range of temperatures for knots E, and F.

\begin{figure*}
\epsscale{1.15}
\plottwo{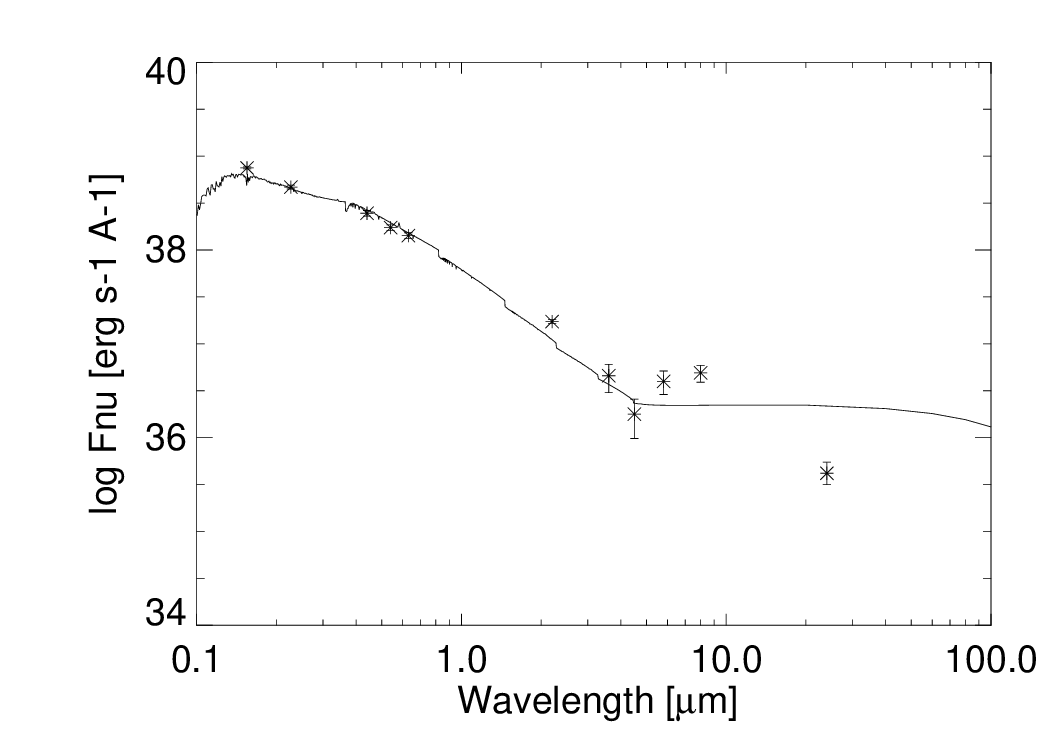}{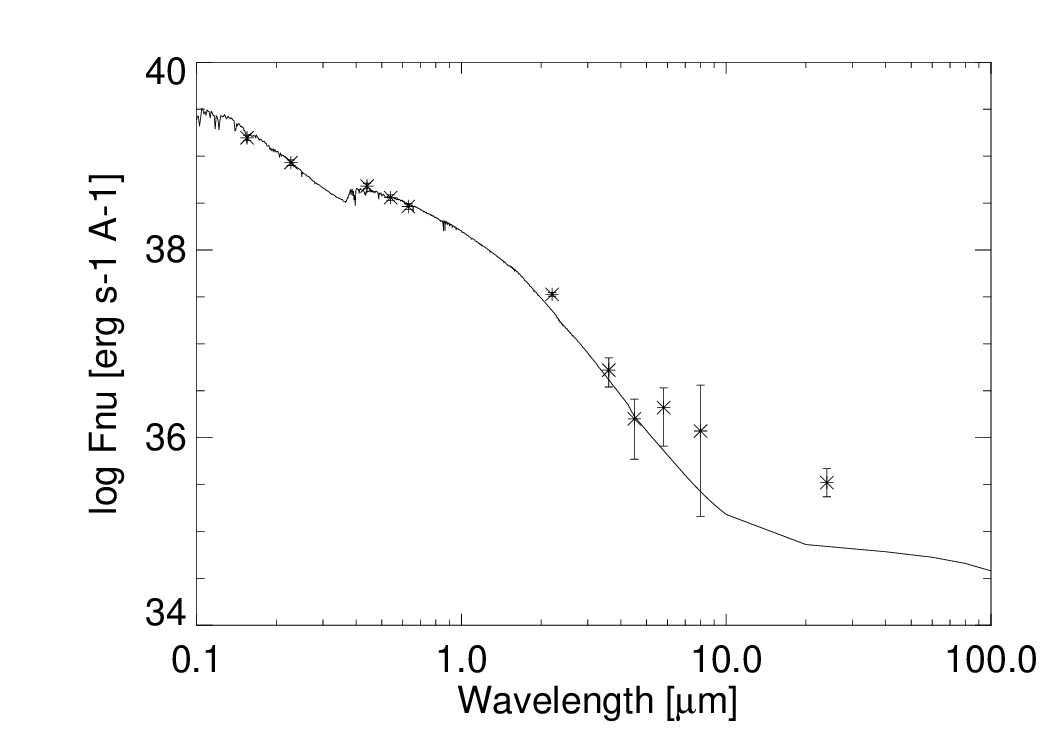}
\caption{Spectral energy distributions for knots C (left) and D (right), with Starburst99 synthetic spectra. The error bars represent photometric measurement errors.}
\label{figsed}
\end{figure*}

In Fig.~\ref{figsed} we show two examples of SED fittings on knots C (with nebular emission) and D (without nebular emission), being good representatives of the two age groups. We can see how the IRAC $5.8\mu$m and $8.0\mu$m fluxes, and the synthetic $24\mu$m flux, deviate from the pure stellar model for knot D, indicating the presence of PAH and thermal dust emission. We can also see that the SEDs are not easily fitted with a simple instantaneous starburst model, and an example is the excess of K-band emission in knot C, observed in Fig.~\ref{figsed} (left). A $K$-band flux excess was also reported in previous studies that fitted SEDs with Starburst99 (\citet{surace99} for warm ULIGs; \citet{mazza92} for the nuclei of Arp 220), which was compensated adding a hot dust (800 K) component contributing to 10-30\% of the emission \citep{surace99}.  

The nucleus of NGC 2445 was modeled using instantaneous (ISF) and continuous star formation (CSF). Continuous star formation is more a reasonable assumption, based on results from ring models, who predict a continuous fuelling of the nuclear starburst through gas inflow. In this case we used a Kroupa IMF with an upper-mass cutoff of 100 M$_\odot$ and a lower mass cutoff of 1 M$_\odot$. We found that the best fit is achieved for the age of 500 Myr. These ages are radically different than those obtained with [NeIII]/[NeII] ratios (5 -- 6 Myr), and this is due to the fact that both [NeII] and [NeIII] lines are produced from the gas excitation by the youngest and hottest massive stars, and that ISF was assumed in the modelling of the lines.

The age determination achieved here is a significant improvement in comparison with past efforts, e. g. \citet{apple92}, as their color-based determinations did not include UV bands. The FUV and NUV, as demonstrated here, are essential to the breaking of the age-extinction-metallicity degeneracy.

\subsection{Star Formation Rates}
\label{sfrs}

With the H$_\alpha$, FUV and mid-infrared continuum fluxes at $15\mu$m, 24$\mu$m, and $30\mu$m we can derive star formation rates for each knot. Using the star formation rate calibration by \citet{kenn98} we can derive the star formation rate from the H$_\alpha$ luminosity, which traces the very young and massive stars. However, the H$_\alpha$ flux is affected by internal extinction, and for the more dusty clusters, we should use the Calzetti calibration law \citep{calzetti07}, which includes the 24$\mu$m luminosity to compensate extinction that affects the H$_\alpha$ emission. Far ultraviolet is also a tracer of young massive stars, but even more affected by extinction than H$_\alpha$. We use here the $A_v$ values in Table~\ref{sed} to correct the FUV fluxes and the calibration by \citet{salim07}, which uses GALEX FUV calibrations. Mid-infrared spectral diagnostics for the SFR were derived in \citet{brandl06}, based on the fluxes at 15$\mu$m and 30$\mu$m. For each spectrum of the knots, we measured the flux at 14.5-15.5$\mu$m and 29.5-30.5$\mu$m, and calculated the SFRs based on the \citet{brandl06} calibrations. In Table~\ref{tabsfr} we list the corrected FUV, H$_\alpha$, 24$\mu$m luminosities, as well as the $15\mu$m and $30\mu$m fluxes. The SFRs derived using each one of these indicators are also listed.

Adding all the SFRs of the knots, we arrive to a total SFR of 2.16$\pm$0.17 M$_{\odot}$yr$^{-1}$ from H$_{\alpha}$ luminosities and 2.40$\pm$0.35 M$_{\odot}$yr$^{-1}$ using H$_{\alpha}+24\mu$m luminosities. This is similar to the 2.5 M$_{\odot}$yr$^{-1}$ derived by \citep{jeske86} using the H$_\beta$ fluxes but below the the far-infrared SFR of 6.21 M$_{\odot}$yr$^{-1}$ calculated from IRAS colors \citep{mohsir90}. It is also far below the SFR derived from the corrected FUV fluxes, which is in total 15.9 M$_{\odot}$yr$^{-1}$.

The SFRs calculated by the \citet{brandl06} method are systematically below the SFR calculated using the \citep{calzetti07} calibrations, which in turn are below the SFRs derived from FUV. This can be explained by the fact that most of the energy is detected in the optical and not absorbed by dust, which makes this system distinct from more archetypal starbursts like M82. Here, the UV and optical radiation from the clusters is heavily absorbed, making the A$_v$ at least 10 times higher than in Arp 143 \citep{beirao08}. This is mitigated in case of the nucleus, which accounts by $\sim80$\% of the total $24\mu$m flux, and thus the discrepancy is smaller. Notice that the star formation rate from UV is heavily dependent on the estimation of A$_v$, which values within a 95 \% confidence can vary by a factor of two, as seen in Table~\ref{sed}. 

\section{The Role of Shocks in the Propagation of Star Formation}





In the past sections we discussed thoroughly the H$_2$ shocks occurring in NGC 2445 and the ages and densities of the knots that compose the star forming ring. 
The strong H$_2$ emission reported this paper exists in and around the density wave, as traced by HI emission shown in \citep{apple92}. This is expected from shocks created as the wave moves out through the disk. This is because the gas, unlike the stars in these models, cannot pass through each other and shocks should develop at caustics. In this section we will discuss the role of these shocks in the development of the knots, and its consequences on the morphology of Arp 143.

\subsection{The evolution of PAH emission with cluster age}
\label{pahage}

The SED modeling in this paper provided a further refinement of the age dating of the knots, which is a main advantage over past attempts \citep{apple92}. It showed that not only the HII region nebulosity but also the clusters themselves are consistent with a recent very young burst.

The evolution of PAH excitation with starburst age has been speculated by other studies, like \citet{roussel05} in their analysis of dust excitation in NGC 300. We can try to observe this behavior taking advantage of our detailed diagnostics of knot ages.

The improvement on the age dating of the knots allowed us to divide them into two groups: those with an age between 2-4 Myr old (knots A, C, E, and F), and those with an age of 7-8 Myr old (knots D and G). What is remarkable is that the younger group of knots corresponds exactly to the knots where PAH counterparts are observed, whereas the older knots are those ``bare'' of PAHs, as discussed in Section~\ref{morphology} and seen in Fig.~\ref{multicolor}. This is the first time this ageing effect in star forming galaxies is observed in such chronological detail.

The lack of PAH emission in older knots means that the PAHs have either ``cleared out'', ceased to be excited, or destroyed during a timescale of 4-5 Myr. The first possibility is that the PAH have been ``swept'' after $\sim6$ Myr by winds from star formation regions. The $8\mu$m shell in knot A, for example, has $\sim3$ kpc in diameter. Assuming a wind velocity of 300 km/s, the shell has been expanding for 5 Myr, a timescale comparable to the age of the knot. Another possibility could be PAH destruction. Small molecules could become more exposed over time to the hard UV field and are subsequently destroyed, as the dust shell expands and the PDRs become exposed to the outer radiation field. However, both these processes also occur in other galaxies where PAHs are observed at later ages, like the Antennae, so the cause for this behavior resides probably on the nature of the ring galaxy system.  
Due to the shock wave traced by the H$_2$ emission, the knots in the ring were formed simultaneously, and therefore we assume an instantaneous burst for all the knots. This means that the UV radiation fades away after $\sim6$ Myr, and no longer excites the dust surrounding the knot. This can be observed in Fig.~\ref{ratios}, where the [NeIII]/[NeII] ratio, a measure of the hardness of the radiation field, decreases dramatically after 6 Myr. The slope of the continuum of knots D and G longwards $25\mu$m indicates the presence of cold dust, meaning that dust is present in these knots, but it is not heated by UV radiation. An older star-forming region with continuous burst like the nucleus of NGC 2445 keeps forming new massive stars which continue to excite PAHs and warm up dust. This could also be the origin of the SFR discrepancy seen in Sect.~\ref{sfrs}, where the SFRs in the knots calculated from the infrared continuum are systematically lower than the SFRs calculated from H$_\alpha$+24$\mu$m luminosities, but not the SFR in the nucleus.

\subsection{The H$_2$ emission front and the simultaneity of knot formation}
\label{h2emission}

The strong H$_2$ emission reported this paper traces a density wave expanding outwards at a constant speed. In response to the density wave, the knots were formed simultaneously in situ, and that is confirmed by the narrow age range found in Table 6. The fact that the H$_2$ follows the shape of the HI overdensity is further confirmation that this is a coherent structure. 

If we assume the shock velocity from \citep{higdon97}, $118\pm30$ km s$^{-1}$ and the distance of the clusters to the nucleus, we can calculate the propagation timescale of the shock from the nucleus outwards. Given that the knots are on an average distance of 10 kpc, the shock wave has been propagating from the nucleus since 85 Myr ago. However, star formation has been occurring in the nucleus way before then, at least since 300 Myr ago. This is consistent with the kinematic picture first advanced in \citet{apple92}: a first interaction with NGC 2444, provoking the onset of star formation in the nucleus 300 Myr ago, and the HI plume observed in \citet{apple87}; and a second interaction $\sim85$ Myr ago, provoking the emergence of a shock wave which creates a gas overdensity visible in HI \citet{apple92,higdon97}, leading ultimately to a ring of young star forming knots. These knots were created almost simultaneously, and as we have seen, blows out surrounding PAHs in a very short timescale, $5-6$ Myr.

Models of ring galaxies have shown the emergence of simultaneous star formation due to an expanding shock wave as a consequence of a head-on collision \citet{gerber96,struck97,lamb00}. However, the timescales and details such as gas distribution, frequency of starbursts, etc., are strongly dependent on the position and kinematics of the galaxies, and there are no models for the particular case of Arp 143.

The future of star formation in the ring of knots will depend on the gas reservoir. Given the mass of atomic and molecular gas, and the SFR for all the knots, we can calculate the duration of star formation in Arp 143 assuming constant SFR. \citet{higdon97} found 1.25$\pm \times 10^9$M$_{\odot}$ of atomic gas in Arp 143, and $2.2\times10^9$M$_{\odot}$ of cold molecular gas. Taking the total gas content and an average SFR of 2.16 M$_{\odot}$, we calculate that the star formation in Arp 143 will last for $\sim2$ Gyr. However, the evolution of Arp 143 as a ring galaxy is also likely to increase its SFR, meaning that this is an upper limit for the duration of star formation.

These estimates do not, however, take into account the dynamical evolution of the system. The intruder galaxy, NGC 2444, is likely to swing back, and disrupt once more NGC 2445. A dynamical model will be extremely useful to predict the future of star formation in this system, and therefore in similar systems at higher redshifts.

\section{Conclusions}

Mid-infrared observations of Arp 143 have been presented in this paper, along with ancillary data including GALEX UV images. Multiwavelength images from UV to mid-infrared show a bright dusty nucleus surrounded by young star forming knots. The four main conclusions of this study are the following:

\begin{itemize}
\item Spectral line maps of the H$_2$ rotational lines show a ridge of warm H$_2$ emission that curves between the nucleus and the western knots. The H$_2$ line flux related to PAH flux is nearly 10 times higher in the ridge than in the nucleus. The flux ratios between the sum of H$_2$ S(0) and S(2) lines over the sum of PAH 7.7$\mu$m and $8.6\mu$m lines reveal that this H$_2$ ridge observed in Fig.~\ref{h2fig} arises from shocked gas behind the wave that provoked the onset of the ring of star forming knots. This is one of the few cases were this kind of feature is seen.

\item With the use of photometry and fitting the SEDs, we improved greatly the age determination of the knots. The knots in the ring are all very young, varying from 2-7.5 Myr old and the nucleus is about 500 Myr old. The ring of knots are a product of a shock wave that has been expanding from the nucleus since $\sim85$ Myr ago, and were formed simultaneously in situ. Behind the shock the molecular gas condenses, and it is traced by the H$_2$ emission ridge. The improvement of age determination is achieved in this study especially due to the GALEX FUV and NUV bands, which are crucial to break the age-extinction-metallicity degeneracy.

\item The distribution of ages of the knots correlate with the presence of PAH emission. Younger 2-4 Myr old knots are associated with PAH emission shells, whereas older 7-8 Myr knots contain little or no PAH emission. The most plausible explanation for this would be an effect of the ageing of the massive stars created in a single instantaneous burst all over the ring, within a timescale of $\sim6$ Myr, after which the UV radiation is no longer energetic enough to excite the PAHs.

\item Given the current reservoir of molecular gas, and assuming that the current star formation rate maintains itself constant, the star formation in Arp 143 will last for about 2 Gyr. However, as it is very likely that the SFR will increase dramatically, this can only be an upper limit for the duration of the starburst. 
\end{itemize}

\acknowledgements
 
We would like to thank Y. Mayya for providing the H$_\alpha$ image. We also thank B. Groves and L. Snijders for making the Starburst99 and Mappings models available. This work is based on observations made with the \textit{Spitzer Space Telescope}, which is operated by the Jet Propulsion Laboratory, California Institute of Technology, under NASA contract 1407. 




\begin{deluxetable*}{ccccccccc}
\tabletypesize{\scriptsize}
\tablewidth{0pt}
\tablecaption{Molecular Hydrogen Line Fluxes (Top Row), Equivalent Widths (Bottom Row) and Diagnostics of the Main Knots\tablenotemark{a}}
\tablehead{
\colhead{Knot} & \colhead{H$_2$ S(3)} & \colhead{H$_2$ S(2)} & \colhead{H$_2$ S(1)} &
           \colhead{H$_2$ S(0)} & \colhead{T (K)\tablenotemark{b}} & \colhead{N$_{H_2}$ \tablenotemark{c}} & \colhead{M$_{H_2}$\tablenotemark{d}} & \colhead{L(H$_2$)\tablenotemark{e}}
}
\startdata
A & & & 0.27 & 0.27 & & & &\\
&&&&&&&&\\
B & 0.22 & 0.31 & 0.39 & 0.18 & & & &\\
&&&&&&&&\\
C & 0.33 & 0.38 & 0.34 & 0.50$\pm$0.14 & & & &\\
& & & & 0.604 & & & &\\
D & 0.39 & 0.38 & 0.27 & 0.20 & & & &\\
&&&&&&&&\\
E & 0.41 & 0.31 & 2.48$\pm$0.03 & 1.12$\pm$0.08 & 137 & 30.6 & 10.8 & 1.35 \\
& & & 0.335 & 0.340 & & & &\\
F & 1.85$\pm$0.10 & 1.17$\pm$0.21 & 4.24$\pm$0.32 & 0.88$\pm$0.12 & 280 & 5.8 & 0.92 & 3.11\\
& 0.386 & 0.183 & 1.106 & 0.194 & & & &\\
G & 0.34 & 0.32 & 4.27$\pm$0.52 & 0.70$\pm$0.12 & 178 & 10.4 & 3.70 & 1.78 \\
& 0.355 & & 1.285 & 0.233 & & & &\\
Ring & 3.37$\pm$0.34 & 1.57$\pm$0.18 & 14.5$\pm$0.6 & 5.70$\pm$0.69 & 236 & 3.1 & 4.45 & 7.46\\
& 0.114 & 0.027 & 0.309 & 0.172 & & & &\\
Nucleus & 1.66$\pm$0.05 & 1.20 & 3.60$\pm$0.15 & 0.91 & 309 & 2.67 & 0.42 & 2.13\\
& 0.022 & & 0.030 & & & & &\\
Nucleus (high resolution) & & 2.72$\pm$0.28 & 5.94$\pm$0.36 & 0.87 & 330 & 1.58 & 0.60 & 3.98\\
& & 0.008 & 0.019 & & & & &\\
\enddata
\tablenotetext{a}{Fluxes in units of $10^{-21}$ W cm$^{-2}$, and the equivalent widths, below the fluxes, are in $\mu$m. Fluxes without errors are upper limits.}
\tablenotetext{b}{Temperatures based on the following slopes: S(0)-S(1) for knot E, S(0)-S(3) for knot F, S(0)-S(1) for knot G, and  S(1)-S(3) for the nucleus}
\tablenotetext{c}{Column densities in units of $10^{19}$ cm$^{-2}$}
\tablenotetext{d}{Warm molecular gas mass in units of 10$^7$ M$_{\odot}$}
\tablenotetext{e}{Luminosities in units of $10^{40}$ erg s$^{-1}$}
\label{tabh2}
\end{deluxetable*}

\begin{deluxetable*}{ccccccc}
\tablewidth{0pt}
\tablecaption{PAH Fluxes\tablenotemark{a} (top row) and Equivalent Widths (bottom row) in the Main Knots}
\tablehead{
\colhead{Knot} & \colhead{6.2 $\mu$m} & \colhead{7.7 $\mu$m} & \colhead{8.6 $\mu$m} &
           \colhead{11.3 $\mu$m} & \colhead{17 $\mu$m complex}  & \colhead{11.3/7.7}
}
\startdata
B & 1.186$\pm$0.395 & 1.717$\pm$0.466 & 0.574$\pm$0.396 & 0.567$\pm$0.256 & 0.146$\pm$0.103 & 0.33$\pm$0.33\\
& 1.14 & 2.49 & 0.949 & 1.20 & 0.045 \\
C & 0.878$\pm$0.229 & 1.952$\pm$0.641 & 0.371$\pm$0.236 & 0.627$\pm$0.128 & & 0.32$\pm$0.26\\
& 1.38 & 3.02 & 0.652 & 1.73 &\\
D & 0.640$\pm$0.359 & 1.397$\pm$0.819 & & 0.446$\pm$0.392 & & 0.32$\pm$0.32\\
& 6.49 & 11.1 & & 2.93 &\\
E & 2.083$\pm$0.814 & 5.824$\pm$2.675 & 1.256$\pm$0.805 & 1.503$\pm$0.455 & 0.665$\pm$0.632 & 0.26$\pm$0.26\\
& 4.42 & 9.03 & 1.72 & 1.74 & 0.756\\
F & 0.922$\pm$0.304 & 4.369$\pm$0.724 & 0.734$\pm$0.334 & 0.972$\pm$0.184 & 0.635$\pm$0.228 & 0.22$\pm$0.10\\
& 2.93 & 9.08 & 1.26 & 1.51 & 1.17\\
G & 0.683$\pm$0.580 & 3.98$\pm$1.449 & 0.724$\pm$0.416 & 0.697$\pm$0.631 & & 0.18$\pm$0.18\\
& 7.89 & 20.7 & 1.46 & 1.25\\
Nucleus & 22.65$\pm$0.441 & 89.52$\pm$0.865 & 13.39$\pm$0.406 & 15.93$\pm$0.312 & 7.346$\pm$0.353 & 0.18$\pm$0.00 \\
& 3.04 & 14.8 & 2.12 & 1.87 & 0.595\\
Ring & 1.14$\pm$0.36 & 4.23$\pm$1.62 & 0.74$\pm$0.34 & 1.04$\pm$0.18 & 0.89$\pm$0.31\\
& 1.87 & 7.97 & 1.38 & 1.60 & 0.94\\
\enddata
\tablenotetext{a}{Fluxes in units of $10^{-20}$ W cm$^{-2}$}
\label{tabpah}
\end{deluxetable*}

\clearpage
\begin{landscape}
\begin{deluxetable}{cccccccccc}
\tabletypesize{\tiny}
\tablewidth{0pt}
\tablecaption{Ionic Line Fluxes and Ratios of the Main Knots}
\tablehead{
\colhead{Knot} & \colhead{[SIV]} & \colhead{[NeII]} & \colhead{[NeIII]} &
           \colhead{[SIII]$18\mu$m} & \colhead{[SIII]$33\mu$m} &
           \colhead{[SiII]} & \colhead{[NeIII]/[NeII]} & \colhead{[SIII]18.7$\mu$m/[SIII]33.5$\mu$m} & \colhead{[SiII]/[SIII]}  
}
\startdata
A & & & 0.53 & & 0.99$\pm$0.09 & 0.87$\pm$0.05 & & & 0.87$\pm$0.15\\
B & 0.24 & & 0.29 & 0.43 & 0.59$\pm$0.10 & 1.16$\pm$0.06 & & & 1.97$\pm$0.51\\
C & 0.36 & & 0.33 & 0.36 & 0.43$\pm$0.12 & 0.87$\pm$0.11 & & & 1.14$\pm$0.90 \\
E & 0.90$\pm$0.04 & 2.25$\pm$0.09 & 1.16$\pm$0.04 & 1.24$\pm$0.27 & 2.48$\pm$0.29 & 2.88$\pm$0.16 & 0.52$\pm$0.04 & 0.50$\pm$0.19 & 1.16$\pm$0.23 \\
F & 0.34$\pm$0.02 & 0.92$\pm$0.05 & 0.62$\pm$0.06 & 0.85$\pm$0.19 & 1.70$\pm$0.22 & 1.64$\pm$0.08 & 0.67$\pm$0.11 & 0.50$\pm$0.20 & 0.96$\pm$0.20\\
G & 0.84 & 0.84 & 0.90 & 1.85$\pm$0.35 & 1.38$\pm$0.21 & 2.05$\pm$0.25 & & 1.34$\pm$0.63 & 1.49$\pm$0.48\\
Nucleus & 0.06 & 30.8$\pm$1.36 & 2.54$\pm$0.08 & 8.71$\pm$0.25 & 14.2$\pm$1.2 & 14.56$\pm$1.60 & 0.08$\pm$0.01 & 0.61$\pm$0.08 & 0.82$\pm$0.22\\
Nucleus (high resolution) & & 63.6$\pm$1.6 & 5.17$\pm$0.60 & 20.7$\pm$0.8 & 28.4$\pm$5.3 & 31.9$\pm$1.6 & 0.08$\pm$0.01 & 0.73$\pm$0.20 & 1.12$\pm$0.21\\
Ring & 1.12$\pm$0.14 & 3.29$\pm$0.38 & 0.99 & 5.21$\pm$0.89 & 10.0$\pm$1.12 & 8.10$\pm$0.68 & & 0.64$\pm$0.15 & 0.81$\pm$0.14\\
\enddata
\tablenotetext{a}{Fluxes in units of 10$^{-21}$ Wcm$^{-2}$}
\label{tablines}
\end{deluxetable}
\clearpage
\end{landscape}

\clearpage
\begin{landscape}
\begin{deluxetable}{cccccccccccccc}
\tiny
\tablecaption{Broadband Fluxes of the Main Knots}
\tablehead{
\colhead{Knot} & \colhead{R.A. (7 46)} & \colhead{Decl. (+39)} & \colhead{FUV\tablenotemark{a}} & \colhead{NUV\tablenotemark{a}} &
           \colhead{B\tablenotemark{a}} & \colhead{V\tablenotemark{a}} &
           \colhead{R\tablenotemark{a}} & \colhead{K\tablenotemark{a}} &
           \colhead{3.6 $\mu$m\tablenotemark{a}} & \colhead{4.5 $\mu$m\tablenotemark{a}} & \colhead{5.8 $\mu$m\tablenotemark{a}} & \colhead{8.0 $\mu$m\tablenotemark{a}} & \colhead{24 $\mu$m\tablenotemark{a}}
}
\startdata
A & 58.21s & $00\min51.24\sec$ & 0.257$\pm$0.005 & 0.311$\pm$0.019 & 0.504$\pm$0.032 & 0.544$\pm$0.035 & 0.587$\pm$0.031 & 0.761$\pm$0.03 & 0.400$\pm$0.172 & 0.257$\pm$0.186 & 0.878$\pm$0.447 & 1.97$\pm$0.74 & 4.87$\pm$1.41 \\
B & 55.38s & $00\min25.75\sec$ & 0.171$\pm$0.003 & 0.231$\pm$0.012 & 3.17$\pm$0.07 & 3.62$\pm$0.08 & 4.00$\pm$0.08 & 6.18$\pm$0.06 & 2.74$\pm$0.32 & 1.73$\pm$0.29 & 2.03$\pm$0.43 & 2.69$\pm$0.65 & 2.62$\pm$0.81\\
C & 54.22s & $00\min21.25\sec$ & 0.093$\pm$0.002 & 0.113$\pm$0.008 & 0.323$\pm$0.026 & 0.360$\pm$0.029 & 0.444$\pm$0.028 & 0.644$\pm$0.02 & 0.501$\pm$0.274 & 0.308$\pm$0.139 & 1.14$\pm$0.32 & 2.71$\pm$0.54 & 2.06$\pm$0.65 \\
D & 53.32s & $00\min18.25\sec$ & 0.197$\pm$0.003 & 0.208$\pm$0.013 & 0.632$\pm$0.040 & 0.755$\pm$0.041 & 0.916$\pm$0.043 & 1.25$\pm$0.03 & 0.587$\pm$0.200 & 0.276$\pm$0.173 & 0.604$\pm$0.368 & 0.645$\pm$0.597 & 1.62$\pm$0.68\\
E & 52.94s & $00\min51.25\sec$ & 0.348$\pm$0.005 & 0.409$\pm$0.019 & 0.754$\pm$0.050 & 0.895$\pm$0.042 & 1.08$\pm$0.04 & 1.39$\pm$0.04 & 0.998$\pm$0.246 & 0.647$\pm$0.337 & 2.78$\pm$0.51 & 6.51$\pm$0.84 & 10.1$\pm$1.4\\
F & 53.97s & $01\min06.25\sec$ & 0.069$\pm$0.002 & 0.077$\pm$0.007 & 0.120$\pm$0.010 & 0.143$\pm$0.012 & 0.173$\pm$0.020 & 0.309$\pm$0.02 & 0.318$\pm$0.142 & 0.193$\pm$0.122 & 1.10$\pm$0.32 & 2.91$\pm$0.55 & 7.50$\pm$0.75\\
G & 54.22s & $01\min18.25\sec$ & 0.167$\pm$0.004 & 0.224$\pm$0.018 & 0.636$\pm$0.047 & 0.808$\pm$0.040 & 0.908$\pm$0.037 & 1.25$\pm$0.03 & 0.560$\pm$ & 0.335$\pm$0.129 & 0.484$\pm$0.279 & 0.841$\pm$0.719 & 5.47$\pm$1.42\\
Nucleus & 55.12s & $00\min55.75\sec$ & 0.049$\pm$0.005 & 0.132$\pm$0.025 & 1.41$\pm$0.053 & 2.48$\pm$0.064 & 3.85$\pm$0.073 & 16.5$\pm$0.1 & 10.7$\pm$0.6 & 7.69$\pm$0.52 & 29.5$\pm$1.0 & 87.6$\pm$1.8 & 173$\pm$7.14\\
\enddata
\tablenotetext{a}{Fluxes in units of mJy}
\tablenotetext{b}{Errors represent $1\sigma$ deviations from the median flux between $22-28\mu$m.}
\label{tabfluxes}
\end{deluxetable}
\clearpage
\end{landscape}

\begin{deluxetable}{cccc}
\tablewidth{0pt}
\tablecaption{Properties of Knots Derived from Starburst99 with Best Fit (Upper Row) and 95 \% Confidence (Bottom Row)}
\tablehead{
\colhead{Knot} & \colhead{Age (Myr)} & \colhead{$A_V$}  & \colhead{Mass (10$^6$ M$_{\odot}$)}
}
\startdata
A & 3.5 & 1.0 & 12.6\\
  & 2-4 & 0.8-1.2 & 10-15.8\\
C & 3.5 & 1.5 & 12.6\\
  & 1-5.5 & 1.1-1.9 & 7.9-15.8\\
D & 7.5 & 0.5 & 12.6\\
  & 6.5-8.5 & 0.4-0.6 & 10-15.8\\
E & 3.5 & 1.2 & 25.1\\
  & 3.5-5 & 1.1-1.4 & 20-25.1\\
F & 2.5 & 1.3 & 6.3\\
  & 0.5-6.5 & 0.8-1.8 & 4.0-10\\
G & 7.5 & 0.6 & 15.8\\
  & 6.5-8.5 & 0.3-0.8 & 12.6-20.0\\
Nucleus (ISF) & 16 & 3.0 & 1122\\
& 15-16 & 2.6-3.4 & 1120-1260\\
Nucleus (CSF) & 500 & 2.7 & 6.3\\
& 400-800 & 2.2-3.1 & 5-7.9 
\enddata
\label{sed}
\end{deluxetable}

\clearpage
\begin{landscape}
\begin{deluxetable*}{cccccccccc}
\tabletypesize{\scriptsize}
\tablewidth{0pt}
\tablecaption{Star Formation Rates of the Main Knots}
\tablehead{
\colhead{Knot} & \colhead{L(FUV)\tablenotemark{a}} & \colhead{L(H$_\alpha$)\tablenotemark{b}} & \colhead{L(24$\mu$m)\tablenotemark{b}} & \colhead{F(15$\mu$m)\tablenotemark{c}} & \colhead{F(30$\mu$m)\tablenotemark{c}} & \colhead{SFR(FUV)\tablenotemark{d}} & \colhead{SFR(H$_\alpha$)\tablenotemark{d}} & \colhead{SFR(H$_\alpha$+24$\mu$m)\tablenotemark{d}} & \colhead{SFR(F(15)+F(30))\tablenotemark{d}}
}
\startdata
A & 1.63$\pm$0.10 & 14.3$\pm$0.2 & 237$\pm$11 & 2.8$\pm$1.3 & 6.7$\pm$1.0 & 1.76$\pm$0.06 & 0.11$\pm$0.02 & 0.11$\pm$0.01 & 0.05 \\	
B & & 8.97$\pm$1.83 & 127$\pm$6 & 1.9$\pm$0.7 & 4.0$\pm$0.7 & & 0.07$\pm$0.02 & 0.07$\pm$0.01 & 0.03 \\
C & 1.87$\pm$0.04 & 16.7$\pm$2.5 & 107$\pm$5 & 1.5$\pm$0.4 & 2.7$\pm$0.7 & 2.02$\pm$0.04 & 0.13$\pm$0.02 & 0.11$\pm$0.01 & 0.02\\
D & 0.390$\pm$0.006 & & & 1.1$\pm$0.1 & 1.6$\pm$0.1 & 0.419$\pm$0.06 & & & 0.012\\
E & 3.49$\pm$0.05 & 45.2$\pm$4.1 & 491$\pm$10 & 4.4$\pm$1.1 & 15.0$\pm$1.6 & 3.77$\pm$0.05 & 0.36$\pm$0.03 & 0.32$\pm$0.02 & 0.10\\
F & 0.880$\pm$0.025 & 27.6$\pm$3.2 & 365$\pm$9 & 3.5$\pm$0.9 & 15.0$\pm$0.8 & 0.951$\pm$0.027 & 0.22$\pm$0.02 & 0.21$\pm$0.01 & 0.10\\
G & 0.421$\pm$0.010 & 12.8$\pm$2.2 & 267$\pm$10 & 2.6$\pm$1.9 & 5.6$\pm$1.2 & 0.463$\pm$0.011 & 0.10$\pm$0.02 & 0.11$\pm$0.01 & 0.04\\	
Nucleus & 6.04$\pm$0.62 & 67.9$\pm$4.96 & 8420$\pm$363 & 62.9$\pm$1.8 & 274.9$\pm$4.8 & 6.53$\pm$0.67 & 0.54$\pm$0.04 & 1.46$\pm$0.27 & 1.63\\
\enddata
\tablenotetext{a}{Luminosities in units of $10^{28}$ erg s$^{-1}$Hz}
\tablenotetext{b}{Luminosities in units of $10^{39}$ erg s$^{-1}$}
\tablenotetext{c}{Fluxes in mJy}
\tablenotetext{d}{Star formation rates in M$_{\odot}$ yr$^{-1}$}
\label{tabsfr}
\end{deluxetable*}
\clearpage
\end{landscape}

\end{document}